\documentclass[twocolumn]{aastex62}
\usepackage{gensymb}
\usepackage{natbib}
\newcommand{\pers}{{\textsc{Perspective}}}

\newcommand{\vkep}{v_\mathrm{kep}}
\newcommand{\vrot}{v_\mathrm{rot}}
\newcommand{\kyr}{{\mathrm{\,kyr}}}

\newcommand{\cm}{{\mathrm{\,cm}}}
\newcommand{\second}{{\mathrm{\,s}}}
\newcommand{\km}{{\mathrm{\,km}}}
\newcommand{\kms}{{\km\second^{-1}}}

\newcommand{\au}{{\mathrm{\,AU}}}

\newcommand{\gm}{{\mathrm{\,g}}}
\newcommand{\gram}{{\gm}}
\newcommand{\massden}{{\gram\cm^{-3}}}

\newcommand{\msun}{{M_\sun}}

\newcommand{\muG}{{\mathrm{\,\ensuremath{\mu}G}}}

\shorttitle{Time evolution with radiative transfer}
\shortauthors{V\"ais\"al\"a et al.}

\begin{document}

\title{Time Evolution of 3D Disk Formation with Misaligned Magnetic Field and Rotation Axes}

\author[0000-0002-8782-4664]{Miikka S. V\"ais\"al\"a}
\affiliation{Academia Sinica, Theoretical Institute for Advanced Research in Astrophysics}
\affiliation{Academia Sinica, Institute of Astronomy and Astrophysics, Taipei, Taiwan}

\author[0000-0001-8385-9838]{Hsien Shang}
\affiliation{Academia Sinica, Theoretical Institute for Advanced Research in Astrophysics}
\affiliation{Academia Sinica, Institute of Astronomy and Astrophysics, Taipei, Taiwan}

\author[0000-0001-5557-5387]{Ruben Krasnopolsky}
\affiliation{Academia Sinica, Theoretical Institute for Advanced Research in Astrophysics}
\affiliation{Academia Sinica, Institute of Astronomy and Astrophysics, Taipei, Taiwan}

\author[0000-0003-4603-7119]{Sheng-Yuan Liu}
\affiliation{Academia Sinica, Theoretical Institute for Advanced Research in Astrophysics}
\affiliation{Academia Sinica, Institute of Astronomy and Astrophysics, Taipei, Taiwan}

\author[0000-0003-3581-1834]{Ka Ho Lam}
\affiliation{Academia Sinica, Institute of Astronomy and Astrophysics, Taipei, Taiwan}
\affiliation{Astronomy Department, University of Virginia, Charlottesville, VA 22904, USA}

\author{Zhi-Yun Li}
\affiliation{Astronomy Department, University of Virginia, Charlottesville, VA 22904, USA}

\email{mvaisala,shang@asiaa.sinica.edu.tw}

\begin{abstract}
Distinguishing diagnostic observational signatures produced by MHD models is essential in understanding the physics for the formation of protostellar disks in the ALMA era. Developing suitable tools along with time evolution will facilitate better identification of diagnostic features.
With a ray-tracing based radiative transfer code \pers\, we explore time evolution of MHD models carried out in \citet{Li2013} --- most of which have $90\degree$ misalignment between the rotational axis and the magnetic field. Four visible object types can be characterized, origins of which are dependent on the initial conditions. Our results show complex spiraling density, velocity and polarization structures. The systems are under constant change, but many of those distinctive features are present already early on, and they grow more visible in time, but most could not be identified from the data without examining their change in time. The results suggest that spiraling pseudodisk structures could function as an effective observation signature of the formation process, and we witness accretion in the disk with eccentric orbits which appear as spiral-like perturbation from simple circular Keplerian orbits. Magnetically aligned polarization appears purely azimuthal in the disk and magnetic field can lead to precession of the disk. 
\end{abstract}


\section{Introduction}

New stars emerge from dense cores within magnetically turbulent molecular clouds, with turbulence originating through a cascade from large scales \citep[mainly driven by supernovae,][]{Korpi1999}. In such a chaotic environment  initial conditions for disk formation can be of many different types. One of which is misalignment of rotation and magnetic fields. The recent results of \textit{Planck satellite} and \textit{Herschel Space Observatory} illustrate this cascade process better from large to small in our Galaxy. The \textit{Planck} measurements of foreground polarization \citep{planckXIX} and their subsequent analysis \citep{planckXXXIII, planckXXXV, Vaisala2018} have shown how galactic magnetic fields shift from diffuse large-scale turbulent field to dense giant molecular cloud (GMC) filaments. While the magnetic field appears to have a coherent GMC-scale structure the polarization fraction is significantly affected by further turbulent fluctuations in the magnetic field \citep{planckXX}. 
A follow up of \textit{Planck} results \citep{C3PO}, the \textit{Herschel Space Observatory} Galactic Cold Cores key project \citep{GCC_I, GCC_II, GCC_III, GCC_IV, GCC_V, GCC_VI, GCC_VII, GCC_VIII, GCC_IX} has demonstrated that cold prestellar and starless cores are highly dynamic structures with diverse features. Therefore, the turbulence cascade has huge influence on protostellar disk formation.
The early models of magnetized collapse \citep[e.g.][]{LiShu1996, Tomisaka2002} assumed highly idealized conditions with simple non-turbulent cores and neatly aligned rotation axis for reasonable practical purposes. However, the simple models are getting more outdated with the complexity revealed by Atacama Large Millimeter/submillimeter Array (ALMA), and star formation theory has to be able to avert so-called magnetic braking catastrophe.

ALMA has provided extensive observations of objects in the early stages of star formation, namely Class 0 stage, which we demonstrate with following examples. \textit{VLA 1623A} is a disk system surrounded by a ring \citep{Sadavoy2018} showing Keplerian-type rotation \citep{Murillo2013}. Indeed, \citet{Harris2018} have found that there might be two components within the inner part of \textit{VLA 1632A}. On the other hand, \textit{HH212} is the classical textbook symmetrical disk seen with an equatorial dust lane and well-collimated outflow ejected from the very center of the disk \citep{Lee2017NatAst, Lee2017SciA, Lee2018}. Yet another example is the \textit{L1448 IRS3B} source, which consists of three protostellar components surrounded by a spiraling density structure \citep{Tobin2016}. In addition, there are objects with X-shaped cavities. Both \textit{TMC1A} \citep{Aso2015, Sakai2016} and \textit{B335} \citep{Yen2018} have rotating X-shaped envelopes. \textit{TMC1A} has been shown to have aligned velocity gradient with the outflow cavity close to the object center \citep{Bjerkeli2016}, whereas \textit{B335} presents coherent magnetization/polarization near the object outflow cavity edges \citep{Maury2018}. The recent observation of IRAS 16293 connect surrounding magnetic field to a protostellar disk \citep{SadavoyIRAS2018, Liu2018}.
The systems so far show significant variation in their structures, and understanding them, and other objects to be uncovered with ALMA, calls for examining the theoretical and observational implications for more than one model or mechanism. In the case of this study, we look at the observational properties of objects formed with magnetic field and rotation misaligned -- one of the proposed solutions to the magnetic braking catastrophe. With such cross-inspection of diagnostic features that can be observed through the input physics, the control process for magnetic braking can be better identified.

In the early days it was assumed that a circumstellar disk would be merely a result of angular momentum conservation of a rotating prestellar core \citep[e.g.][]{Bodenheimer1995}. However, magnetic field proved to make the matter more complicated. 
Magnetic braking catastrophe results from accumulation of magnetic flux into the collapse center, in ideal-MHD conditions \citep[See][for more extended reviews]{Lizano2015, LiPPVI}. When a core collapses, magnetic field converges towards the center growing in strength, and during the process the field becomes strong enough to oppose any azimuthal movement \citep{Allen2003, Galli2006}. This will also lead to overly efficient removal of angular momentum by the outflow, inhibiting disk formation  \citep{Mellon2008, Hennebelle2008}. Magnetic braking can also lead to interchange instabilities, disturbing its ordered structure: when the magnetic field is decoupled from the central protostar, some regions around the protostar can grow strong enough to oppose free-fall motion \citep{LiMcKee1996}. This phenomenon can produce a system fragmented by so-called decoupling enabled magnetic structures \citep[DEMS,][]{Zhao2011, Ruben2012}, which appear as low-density, high-magnetic field holes around the protostar, which block the rotation of the accretion flow. However, because we know that magnetic protostellar disks exist \citep{Williams2011}, something has to make disk formation possible despite the magnetic field.

There are various suggested mechanisms for resolving magnetic braking catastrophe. They are not contradictory, because in a complete, realistic system all of them could be present in some proportion. However, it is interesting to ask which mechanism dominates and under what conditions. One group of attempted solutions are the non-ideal MHD effects of Ohmic resistivity \citep{Dapp2010, Ruben2010, Machida2014}, ambipolar diffusion \citep{KK2002, MellonLi2009, Li2011} and Hall-effect \citep{Ruben2011, Tsukamoto2017}. Non-ideal MHD effects, as long as they are strong enough, would loosen the grip of matter on the magnetic field, thus preventing excessive magnetic braking. However,  \citet{Ruben2012} have demonstrated that non-ideal effects by themselves might not prevent the catastrophe, at least when the canonical level of ambipolar diffusion is adopted.

On the other hand, turbulent diffusion and reconnection of magnetic fields could also make efficient disk formation possible \citep{SantosLima2012, Joos2013, Li2014, Matsumoto2017, Kuffmeier2017, Gray2018}. The turbulence studies have provided promising results. Turbulence could create direct diffusion of the magnetic field --- particularly in the form of reconnection \citep{LazarianVishniac1999}. Additionally, the turbulent cascade from the large scales could affect the alignment of the magnetic field and the inflow during star formation process, making it more beneficial for disk formation \citep{Kuffmeier2017,Gray2018}: this point also connects to the third option, the misalignment of the rotation axis with respect to the magnetic field, which is the specific focus of this study. 

There have been a number of studies examining the misalignment \citep[such as][]{Machida2006, Hennebelle2009, Joos2012}.  These studies had shown that with a mass-to-flux ratio as low as $\lambda \gtrsim 3$ misaligned collapse could produce a rotationally supported disk (RSD). However, a later study by \citet{Li2013} (hereafter LKS13) argued that $\lambda \gtrsim 4$ would be minimum to allow for misaligned disk formation, as their simulation starts from a uniform spherical cloud, whereas earlier models started with a density profile condensed at the center. Therefore, \citet{LiPPVI} considered that it would be rare that collapse would have both a misaligned magnetic field and $\lambda \gtrsim 4$, suspecting that a more hybrid approach would be needed to fully resolve the magnetic braking catastrophe. 

Since the publication of LKS13, \citet{Gray2018} have explored misalignment with a power-law scaled initial turbulence starting from a spherical $300$\,$\msun$ molecular cloud. They tested two implementations of turbulence: one had the total angular momentum of the turbulence aligned to the magnetic field, the other had it misaligned. They found that aligned turbulence, while leading to efficient star formation, could not produce disks, regardless of the energy of the initial turbulence or levels of mesh refinement. On the other hand, the misaligned turbulence would make disk formation possible, with large-scale cloud rotation/magnetic field misalignment leading to even more star formation. Therefore, \citet{Gray2018} argue that turbulent diffusion, just by itself, cannot fully account for disk formation. They suggest that \textit{turbulent misalignment} is an essential element of the process. This means turbulence would, by affecting fluid movements and magnetic field alignment locally,  lead to magnetic field alignments over time which would favor disk formation.

In the light of the results of \citet{Gray2018}, the role of misalignment could be seen as a part of the process related to turbulent collapse at molecular cloud scales. In such a scenario, we will likely find at least some cores for which rotation axis and larger-scale magnetic fields are almost orthogonal to each other --- aligned so due to the influences of larger scale turbulence. Therefore, it is meaningful to build the basis for observational understanding of such object types. 

The series of MHD models, results of which are used as the basis of radiative transfer modeling in this study, were in part presented in LKS13. Of special interest are models G and H which show signs of disk formation. We use these models to gain insights concerning the observational signatures of misaligned disk formation process via radiative transfer modeling with key diagnostic elements. We will show that, such signatures can be found, and indirect signatures might tell more about the magnetic field than the linear polarization. Also new understanding of the original MHD models themselves can be gained. To achieve our aims we used the simple but efficient radiative transfer code \pers\ developed within the group to estimate the polarized continuum emission from our model disks along with column density information and position velocity diagrams.

The rest of the paper is structured as follows. First, we describe the MHD model of LKS13 in Section \ref{sec:mhdmodel}. In Section \ref{sec:methods} we describe the basic functionalities of the radiative transfer code \pers\ and other analysis and fitting methods we have used. In Section \ref{sec:results} we describe general results of the radiative transfer modeling based on LKS13, and how we categorize the results by their nature. In Section \ref{sec:disc} we discuss what the best observational signatures produced may be, and how they are related to the physics and kinematics. We argue that VLA 1623A \citep{Murillo2013,Murillo2013SMA, Sadavoy2018, Harris2018} could be a potential candidate for the misaligned disk formation scenario. Section \ref{sec:conclusions} summarizes the main results.

\section{MHD Models}\label{sec:mhdmodel}

\begin{table}
\begin{center}
\caption{Models of LKS13 and their corresponding visual types. \label{tbl:types}}
\begin{tabular}{rllllll}
\tableline
\tableline
Model & $\lambda$ & $\theta_0$ & $\eta$ & $M_*$ & Visual type  \\
 & & & ($10^{17}$ cm$^2$ s$^{−1}$) &	($M_\odot$)      &       \\
\tableline
\textbf{A} & $9.72$ & $ 0\degree$ & $1  $ & $0.24$ & LP \\ 
B          & $4.86$ & $ 0\degree$ & $1  $ & $0.22$ & LP \\
C          & $2.92$ & $ 0\degree$ & $1  $ & $0.33$ & LP \\
\tableline
D          & $9.72$ & $45\degree$ & $1  $ & $0.21$ & LP/LS \\
E          & $4.86$ & $45\degree$ & $1  $ & $0.35$ & LP \\
F          & $2.92$ & $45\degree$ & $1  $ & $0.27$ & LP \\
\tableline
\textbf{G} & $9.72$ & $90\degree$ & $1  $ & $0.38$ & CS \\
\textbf{H} & $4.86$ & $90\degree$ & $1  $ & $0.46$ & LS \\
\textbf{I} & $2.92$ & $90\degree$ & $1  $ & $0.47$ & LA \\
\tableline
M          & $9.72$ & $90\degree$ & $0  $ & $0.10$ & CS \\
N          & $9.72$ & $90\degree$ & $0.1$ & $0.26$ & CS \\
\tableline
P          & $4.03$ & $90\degree$ & $1  $ & $0.32$ & LS \\
Q          & $3.44$ & $90\degree$ & $1  $ & $0.22$ & LA \\
\tableline
X          & $4.00$ & $90\degree$ & $1  $ & $0.31$ & LA \\
Y          & $5.00$ & $90\degree$ & $1  $ & $0.33$ & LA/LS \\ 
Z          & $6.00$ & $90\degree$ & $1  $ & $0.30$ & LS \\
\tableline
\multicolumn{7}{l}{\textbf{Notes.} Clear Spiral (CS), Leaking Spiral (LS),  }\\
\multicolumn{7}{l}{Looped Axis (LA), Looped Plane (LP). Boldface highlights  }\\
\multicolumn{7}{l}{the representative model cases featured in the figures.}
\end{tabular} 
\end{center}
\end{table}

The work presented in this paper has been built on the simulations conducted by LKS13, which aimed to study effects of magnetic field misalignment, particularly in comparison with \citet{Joos2012}. LKS13 used non-uniform spherical grid with a resolution of $96 \times 64 \times 60$ with inner and outer radial boundaries at $10^{14}\cm \approx 6.7\au$ and $10^{17}\cm \approx 6700\au$ respectively. The radial cell size is most resolved near the inner boundary, where $\Delta r=5\times 10^{12}\cm \approx 0.3\au$. The $\theta$-grid is also non-uniform, with higher resolution towards the equator ($\sim 0.63\degree$), and lower towards the polar axis ($7.5\degree$).
For boundary conditions, at radial boundary, standard outflow conditions were imposed. The azimuthal direction is naturally periodic, and reflective boundaries were imposed at the two polar axes. The mass flowing out of the grid in the inner radial direction accretes to a central point mass.

LKS13 ran the MHD models listed in Table \ref{tbl:types}. The models started from a sphere with uniform density of $\rho_0 = 4.77 \times 10^{-19}\massden$, radius of $R=10^{17}\cm \sim 6700\au$ and solid-body rotation of $\Omega_0 = 10^{-13}\second^{-1}$. Their varied parameters included the mass-to-flux ratio $\lambda$, the angle between rotation axis and the magnetic field $\theta_0$ and degree of Ohmic resistivity $\eta$. The orientation and strength of the magnetic field was adjusted to get the intended $\lambda$ and $\theta_0$ with initial field strengths $B_0 = 10.6$, $21.3$ and $35.4\muG$. Models P and Q were set to estimate the limit value for $\lambda$ being able or not to produce a rotationally supported disk. Models X, Y and Z had a magnetic field proportional to the column density along each field line. 

All models started from ideal-MHD, but an explicit small and spatially uniform Ohmic resistivity was added after the initial collapse, to avoid numerical instabilities. LKS13 state that added resistivity has little influence on the essential dynamics of their model. The system was governed by a barotropic equation of state, isothermal at low densities, and polytropic at high. Self-gravity was included, and as well as the gravity due to the accreting central point mass.

LKS13 found that the lack of a strong outflow allows the misaligned case to conserve more of the angular momentum, leading to disk formation, unless the system had too strong a magnetic field with $\lambda < 4$. They found that misaligned collapse would be magnetically dominated with rotation approaching Keplerian, and inflow was slower than the free-fall velocity. In the process, the magnetic field would wrap itself into a spiraling shape, with adjacent spirals having opposing magnetic polarities. 

With aligned collapse or misaligned case with strong magnetic field, the the concentration of magnetic flux near the accreting protostar would lead to the formation of DEMS \citep{Zhao2011, Ruben2012}. DEMS form when the magnetic pressure around the central object grows strong enough to oppose the ram pressure from the collapse, which will result in areas with low density. The edges surrounding the DEMS, on the other hand, have enhanced densities, appearing then as loop-like structures. 

LKS13 concluded that magnetic misalignment would not be able to explain all disk formation, because according to their estimate, probability for having favorable conditions for misaligned disk formation would be $\sim 12\%$ or less. Likelihood of suitable condition would be rare. 
Such estimate could be, however, too pessimistic. As argued earlier, early protostellar envelopes can have multiple types, for some of which misaligned formation can be relevant.  Large-scale turbulence could result in exotic smaller scale orientations, and \citet{Kuffmeier2017} argue that the mass-to-flux ratio would not be a hard determining feature for the possibility of circumstellar disk formation. In addition, new model results by \citet{Gray2018} demonstrate that magnetic field misalignment can be an important part of the collapse process to some degree.  Therefore, modeling observable features of misaligned magnetic field would still be called for so that such behavior could be identified in nature. 

\section{Methods} \label{sec:methods}

MHD models such as those of LKS13 are three-dimensional dynamical entities and as such their properties can only be fully understood by analysis techniques which take this into account. While LKS13 did detailed analysis of their physics, apart from some measured quantities, their analysis relied on 2D cutout slices, which as will be discussed later, miss out some essential perspectives on how the system functions in three dimensions and their respective interpretations.

The old analysis could address the observational implication only in a limited way, which is a matter that requires extended consideration. Therefore, to deepen the original work we aimed to examine the LKS13 data like observable objects. This would mean doing simple radiative transfer modeling of column density, polarization and projected physical quantities like velocity --- with a large collection of MHD model frames. 

A simple ray-tracing code \pers\ was built for such a purpose, and this is the first time that the code has been used in a scientific study.
\pers\ is capable of integrating the above listed quantities from a chosen viewing angle producing estimates for observations. Due to its flexibility and efficiency it can be also treated as a data viewing tool. In this section we introduce its basic functioning and further data-analysis methods. 
\begin{figure*}
\plotone{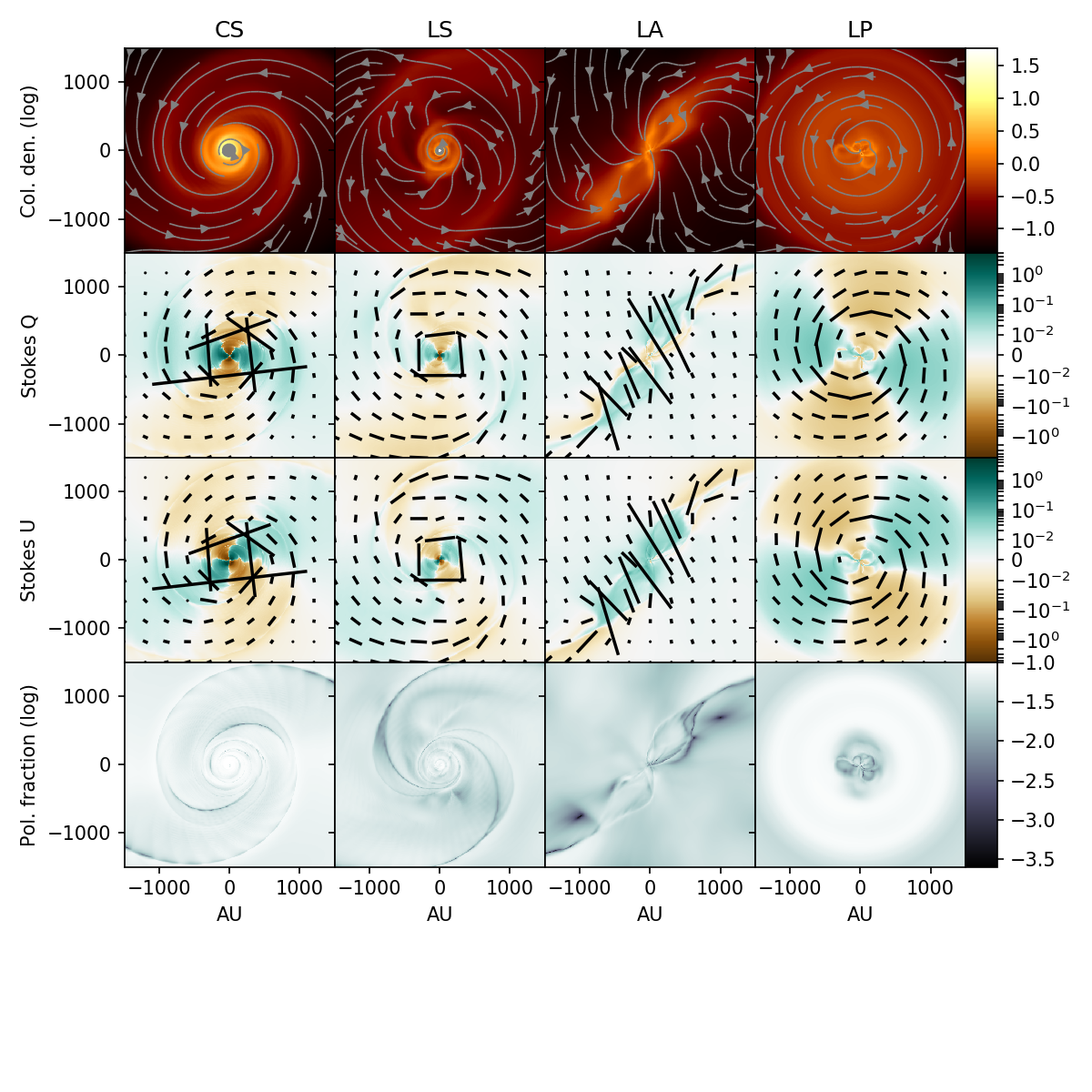}
\caption{Models G (\textit{CS}-type), H (\textit{LS}-type), I (\textit{LA}-type) and A (\textit{LP}-type) viewed from above. The streamlines on the column density maps show the direction of flows along the picture frame. The black lines over Stokes Q and U parameters depict the direction of magnetic field estimated from the polarization angles (``B-vectors''). (Large-scale) \label{fig:all_above}}
\end{figure*}

\begin{figure*}
\plotone{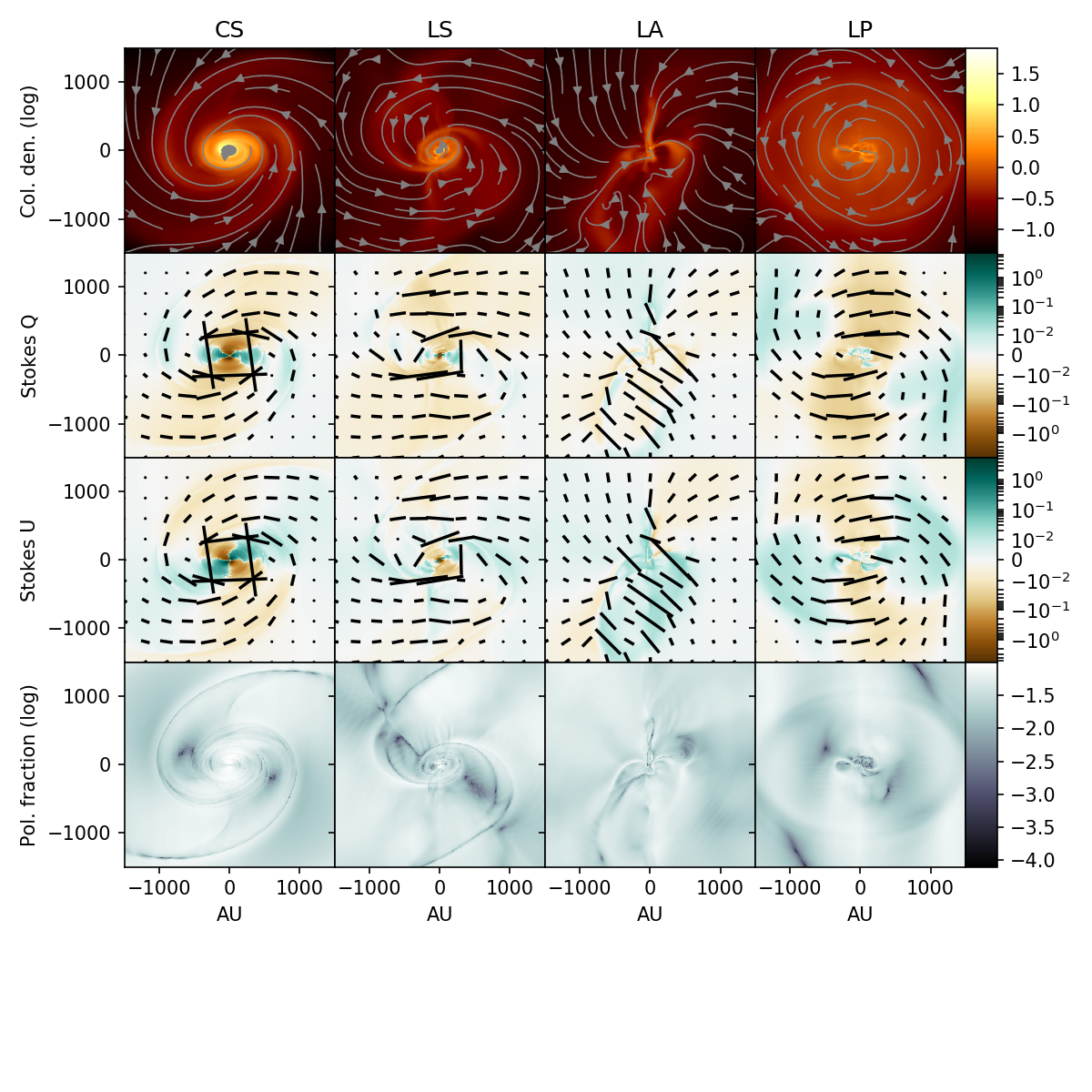}
\caption{Models G (\textit{CS}-type), H (\textit{LS}-type), I (\textit{LA}-type) and A (\textit{LP}-type) viewed from $45\degree$ angle. (Large-scale) \label{fig:all_45}}
\end{figure*}

\begin{figure*}
\plotone{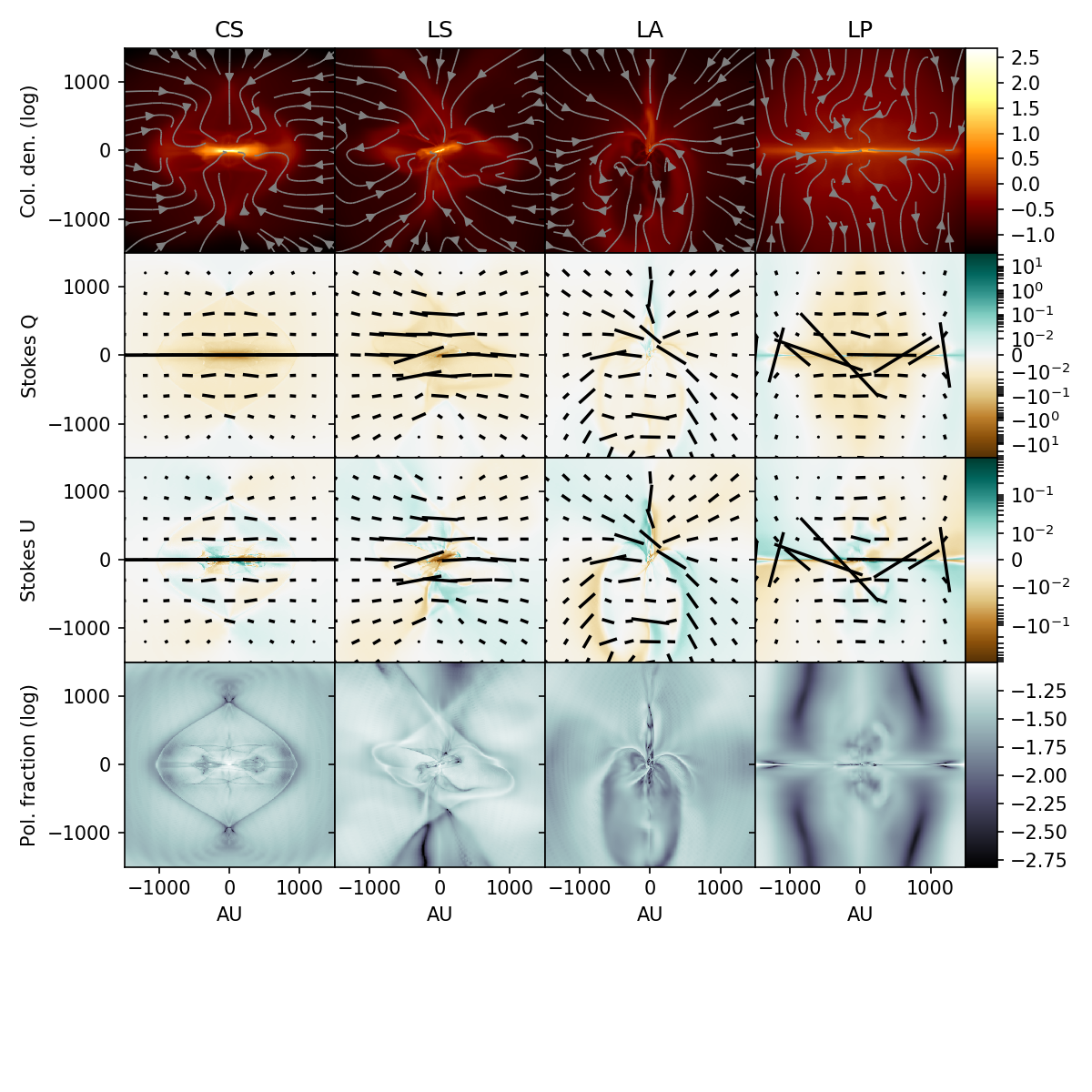}
\caption{Models G (\textit{CS}-type), H (\textit{LS}-type), I (\textit{LA}-type) and A (\textit{LP}-type) viewed edge-on. (Large-scale) \label{fig:all_edgeon}}
\end{figure*}

\subsection{Ray tracing and Polarization}

\pers\ calculates radiative transfer with a ray tracing method. To do so, \pers\ pre-calculates all paths of the rays across the examined computation volume, performing relevant interpolations along the way, based on the ZeusTW data cube and a choice of viewing angles and zoom levels. After determining and interpolating the rays the gas column densities can be integrated along the line of sight (LOS) with 
\begin{equation}
    \Sigma = \int \rho\,ds
\end{equation}
where $\rho$ is the gas density, $ds$ the LOS length differential and the integration is computed along the whole length of the ray. 

\pers\ is able to calculate the Stokes parameters of emitted polarization based on the method presented by \citet{Fiege2000}, assuming an optically thin medium. Therefore, the polarization parameters $q$ and $u$ are obtained by
integrating the rays along the LOS as
\begin{equation}
    q = \int \rho\,\cos\!2\psi\,\cos^2\!\gamma\,ds
\end{equation}
\begin{equation}
    u = \int \rho\,\sin 2\psi\,\cos^2\!\gamma\,ds,
\end{equation}
where $\gamma$ is the angle between the plane of the sky (POS) and the local direction of the magnetic field, and $\psi$ measures the direction of the local polarization angle tilted $90\degree$ from the projected POS magnetic field direction within the given integration element.

In addition, we get the magnetic structure parameter 
\begin{equation}
    \Sigma_2 = \int \rho \left(\frac{\cos^2 \gamma}{2} -\frac{1}{3} \right)\, ds.
\end{equation}
From these an estimate of polarization fraction $p$ can be calculated by using, 
\begin{equation}
    p = \langle \alpha \rangle \frac{\sqrt{q^2 + u^2}}{\Sigma - \langle \alpha \rangle \Sigma_2}
\end{equation}
where we assume $\langle \alpha \rangle = 0.1$ as in \citet{Fiege2000}. This is an approximate observationally estimated scaling factor, and our analysis will avoid relying on the absolute values of $p$, and focus more on its variation. Stokes $I$, $Q$ and $U$ are presented in computer units for the sake of simplicity, as in our analysis only relative values have meaning. The scaling of the units assumes a single dust species and a constant dust temperature with related coefficients scaled to unity. In addition, dust alignment efficiency and gas-to-dust relation are kept constant. See \citet{Fiege2000} for the use of this model under wider assumptions.

While the method is simple, it is efficient when calculating Stokes parameters from all of the hundreds of MHD model frames from multiple perspectives. The method connects polarization directly with the magnetic field geometry. However, our approach does not account for other mechanisms affecting polarized emission, such as self-scattering from thermal emission \citep{Kataoka2015, Yang2016} or if dust grains are directly aligned by the radiative flux from the central protostar itself \citep{Tazaki2017}. In addition, we are not performing our analysis with multiple dust species. Therefore, our polarization results are related only to object's density and magnetic field geometry. This is a practical approach because magnetic alignment is straightforward to compute with \pers\ method, while other mechanisms would require more complex, and therefore more computationally heavy approach, suitable only to a small number of MHD model frames. Because there is not yet complete consensus on the impact of all these polarization mechanisms, estimates of magnetic alignment effect can still be very informative to understand its potential role.

In addition to the polarized emission, we have calculated other derived values described in detail in Appendix \ref{sec:velgauss}. These include line of sight velocity averages for all Cartesian components $v_\mathrm{LOS}$, $v_\mathrm{W}$ and $v_\mathrm{N}$ --- or line of sight, west and north directional velocities respectively. Also disk height $H_{z}$, disk tilt angle $\xi$, apparent momentum $p_\mathrm{mom}$ and average peak location $\zeta$ are calculated based on a Gaussian fit.

\begin{figure*}
\plotone{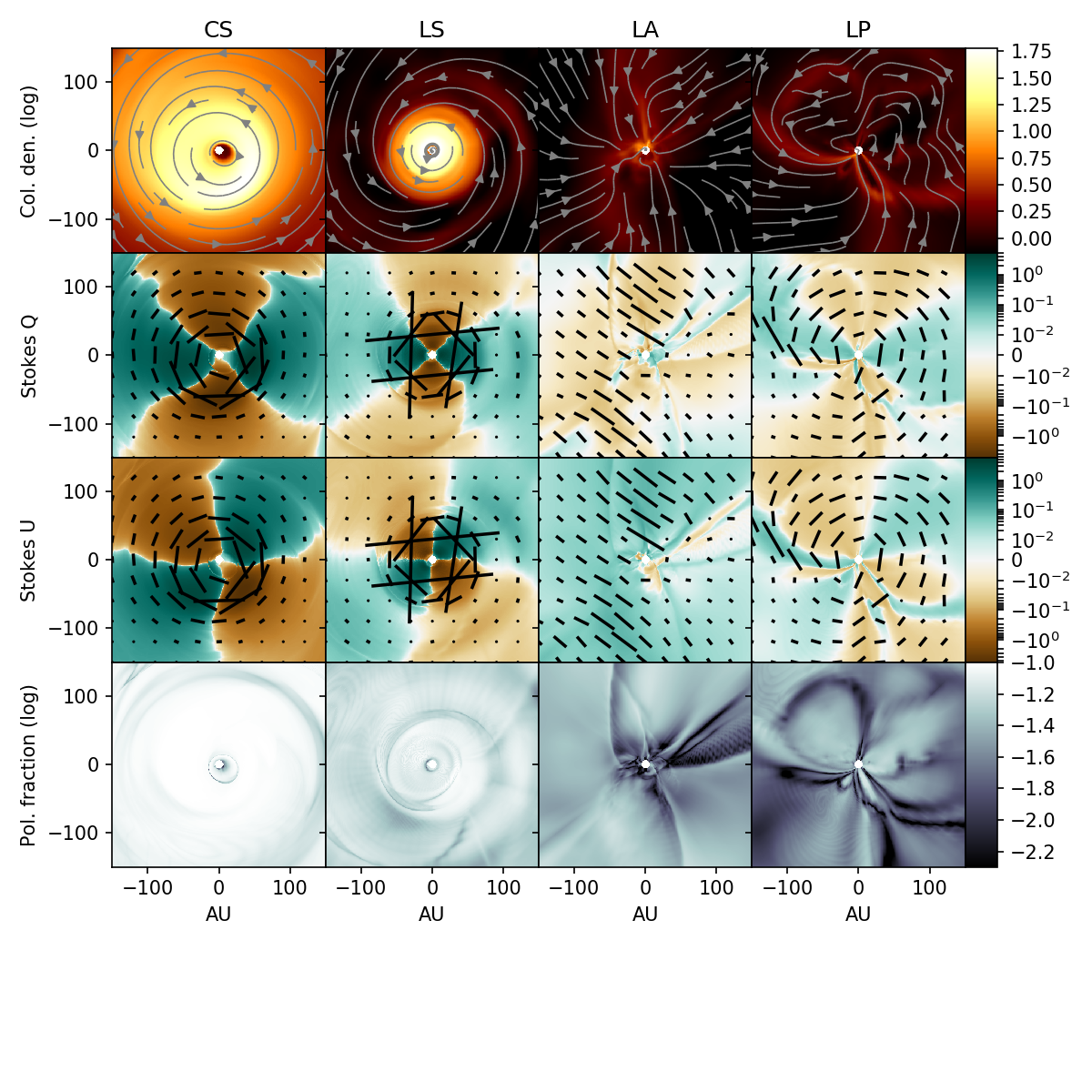}
\caption{Models G (\textit{CS}-type), H (\textit{LS}-type), I (\textit{LA}-type) and A (\textit{LP}-type) viewed from above. (Small-scale) \label{fig:close_above}}
\end{figure*}

\begin{figure*}
\plotone{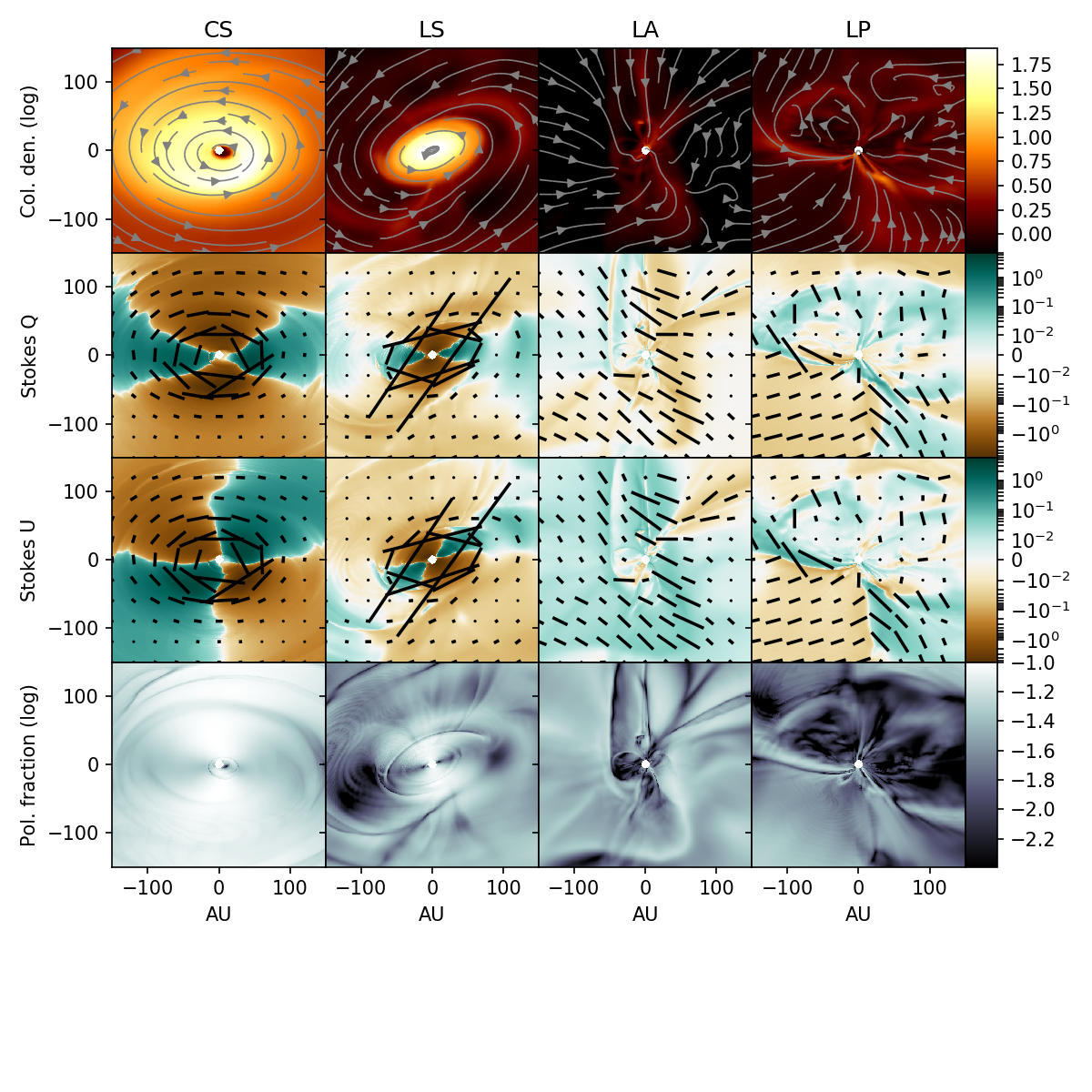}
\caption{Models G (\textit{CS}-type), H (\textit{LS}-type), I (\textit{LA}-type) and A (\textit{LP}-type) viewed from $45\degree$ angle. (Small-scale)\label{fig:close_45}}
\end{figure*}

\begin{figure*}
\plotone{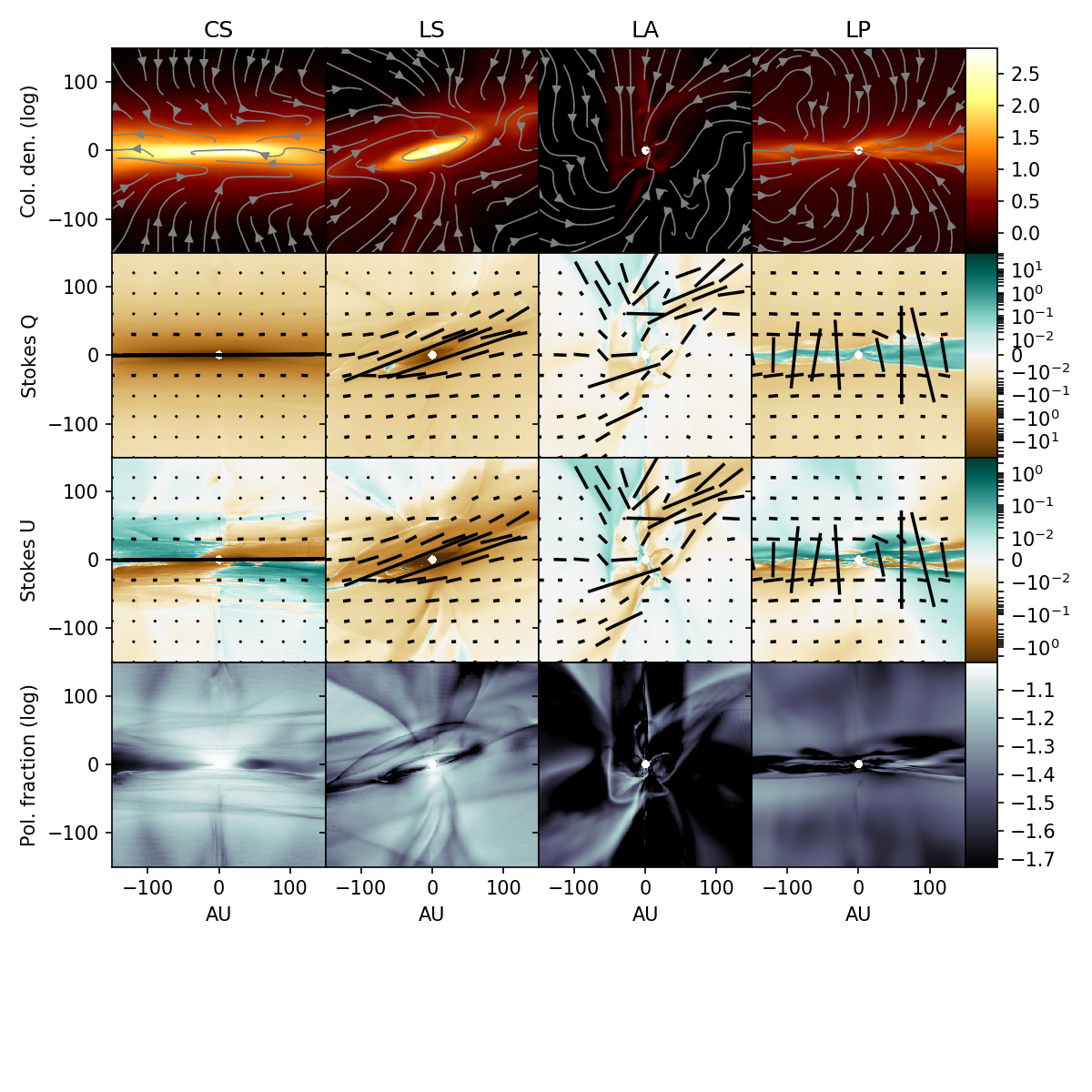}
\caption{Models G (\textit{CS}-type), H (\textit{LS}-type), I (\textit{LA}-type) and A (\textit{LP}-type) viewed edge-on. (Small-scale)\label{fig:close_edgeon}}
\end{figure*}

\section{Results} \label{sec:results} 

With the methods presented above, we produced hundreds of images. Because it would be impossible to show all of them, we have to focus on highlights and generalized descriptions. Fortunately, we have found that certain basic elements stay fairly consistent and we can present a simplified four-type categorization for majority of the results. 

For the results to be helpful for observational interpretation, there is a need to be able to clearly differentiate the types of objects produced by the MHD modeling in reasonable categories that are based on their visual nature. Looking that way, essentially four different special object types can be seen from the results. These fairly well represent what LKS13 described as ``robust'' and ``porous'' rotationally supported disks (RSD), but aim to be more descriptive from an observational standpoint.

These categories are arbitrarily named as \textit{Clear Spiral}, \textit{Leaking Spiral}, \textit{Looped Axis} and \textit{Looped Plane}: 
\begin{enumerate}
    \item \textit{Clear Spiral (CS)} shows a distinctive spiral-like shape, which is both visible in the column densities and the observed polarization. However, no outflows are visible, and no axial infall streams can be seen from column densities.
    
    \item \textit{Leaking Spiral (LS)} is similar to the \textit{CS} but the system shows clear signs of precession of the disk plane and of the axial infall stream (``infunnel'') is visible as a cone-like shape in column densities, with some sporadic outflows. Both \textit{CS} and \textit{LS} cases show two-armed spiral-like pseudodisk. 
    
    \item \textit{Looped Axis (LA)} type object shows horizontal, growing loop-like features identifies as DEMS with magnetic braking preventing disk formation. An arched bar-shaped pseudodisk is visible face-on. 

    \item \textit{Looped Plane (LP)} is a system which forms DEMS loops near the rotational plane, with magnetic braking preventing disk formation. There is a flat pseudodisk but no disk.  
\end{enumerate}
For visual samples of these object types, see Figures \ref{fig:all_above}, \ref{fig:all_45}, \ref{fig:all_edgeon}, \ref{fig:close_above}, \ref{fig:close_45} and \ref{fig:close_edgeon}. Table \ref{tbl:types} groups the physical models with the visual types. These essentially contain all of the produced object types in a clear-cut manner. Exceptions would be Models D and Y, which very much resemble \textit{LS} case, but are weaker than others. 

Model D resembles Model A in many respects, as DEMS are clearly visible in its rotational plane, lacking the coherent disk-like surface structure of Models G and H, however resembling them in the outer parts of the system. However, Model D is more disk-like than Model A, and LKS13 classified it as a ``porous'' disk. Because \pers\ results do not present a convincing and persistent disk, we have decided not to give it a special category in our discussion. Regardless, a particular feature of Model D is its surrounding pseudodisk, oriented along the $45\degree$ angle of the magnetic field. Model Y, classified as ``porous disk'' by LKS13, resembles an LA case, but shows a weak indication of a disk-like structure with no precession, making it another boundary case in the set of models.

Within the models under the stated categories there can still be physical differences. However, the differences are so fine that they do not produce distinctively different observational features. Therefore, most of the time we focus on the representative case for each category, which are Models \textbf{G}, \textbf{H}, \textbf{I} and \textbf{A} respectively (Table \ref{tbl:types}). 

The four categories are related to their characteristic initial conditions, which can be reduced to the relative orientation of the magnetic field and its strength. Table \ref{tbl:initcond} lists the basic initial conditions leading to each visual type. The `weak field' refers to mass-to-flux ratio ($\lambda$) of $9.72$, in LKS13. The `medium field' corresponds to $\lambda = 4.03$ -- $4.86$ and the `strong field' to $\lambda = 2.92$ -- $3.44$, if we list only misaligned collapse cases which started from an initially uniform magnetic field.

\begin{table}
\begin{center}
\caption{Reference table for corresponding visual types with initial conditions.\label{tbl:initcond}}
\begin{tabular}{rl}
\tableline
\tableline
Visual type & Initial conditions \\
\tableline
Clear Spiral (CS)        & Misaligned weak field         \\  
Leaking Spiral (LS)      & Misaligned medium field       \\
Looped Axis (LA) & Misaligned strong field       \\
Looped Plane (LP)        & Aligned field of any strength \\
\tableline
\end{tabular}

\end{center}
\end{table}

The misalignment of rotation and magnetic field has a significant dynamical importance, beyond the strength of the field itself. 
With the misaligned magnetic field, the strength of the field itself is of dynamical importance. While the strong case prevents it, the medium and weak fields lead to disk formation. 

When examining these objects, two approximate scales are helpful. $R \sim 1000$\,AU scales give a sense of the whole model object. It especially highlights the features relating to the surrounding envelope, as some of the particularities relate to the infall from the envelope. On the other hand, $R \sim 100$\,AU looks into the inner part where the \textit{CS} and \textit{LS} cases form an early disk. Let us look first at large scales. 

\subsection{Large scales ($R \sim 1000$\,AU)}\label{sec:large}

In \textit{CS} case, by the end of the MHD simulation, viewed from above, the system has formed a two-armed spiral-like shape that is relatively regular (see Figure \ref{fig:all_above}), although the inner region is more tangled than the outer region. This appearance can be still perceived from $45\degree$ inclination (see Figure \ref{fig:all_45}). 
Looking edge-on, there are no visible outflows, and the inner disk is surrounded be an almost symmetric shell (see Figure \ref{fig:all_edgeon}). 
However, the velocity analysis shows that there are inflows in both polar and plane-wise direction, discussed more in Section \ref{sec:velocity}. The inflows follow the spiral ridges. 

\begin{figure*}
\plotone{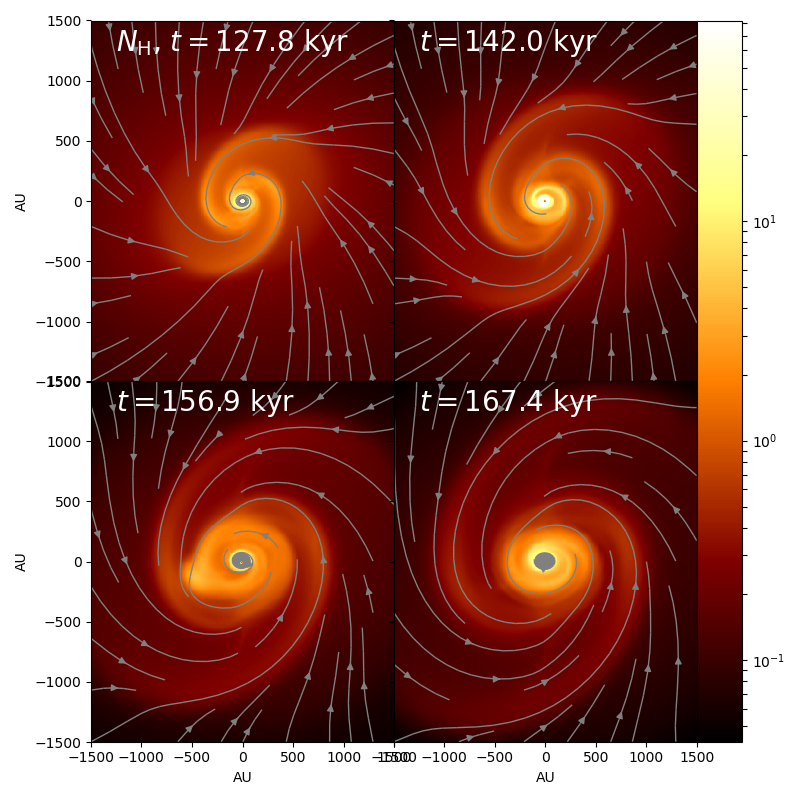}
\caption{Time development the Model G (\textit{CS} case), viewed from $45\degree$ angle. The colors refer to column densities. \label{fig:timeframeCS}}
\end{figure*}

Dynamically the outer and inner spirals of the \textit{CS} case behave differently. The shape of the outer spiral stays relatively stable in time, but the rapidly rotating inner spiral shows oscillatory modes where spiral arms are no longer clearly distinct. In later frames of the Model G simulation, we see that a blob on that inner spiral becomes super-Keplerian, and momentarily finds a higher orbit along the rotational plane before rejoining the spiral inflow. This is shown in the third panel of Figure \ref{fig:timeframeCS}, and the Gaussian fit analysis of the oscillations is presented in Section \ref{sec:ts}.

\begin{figure*}
\plotone{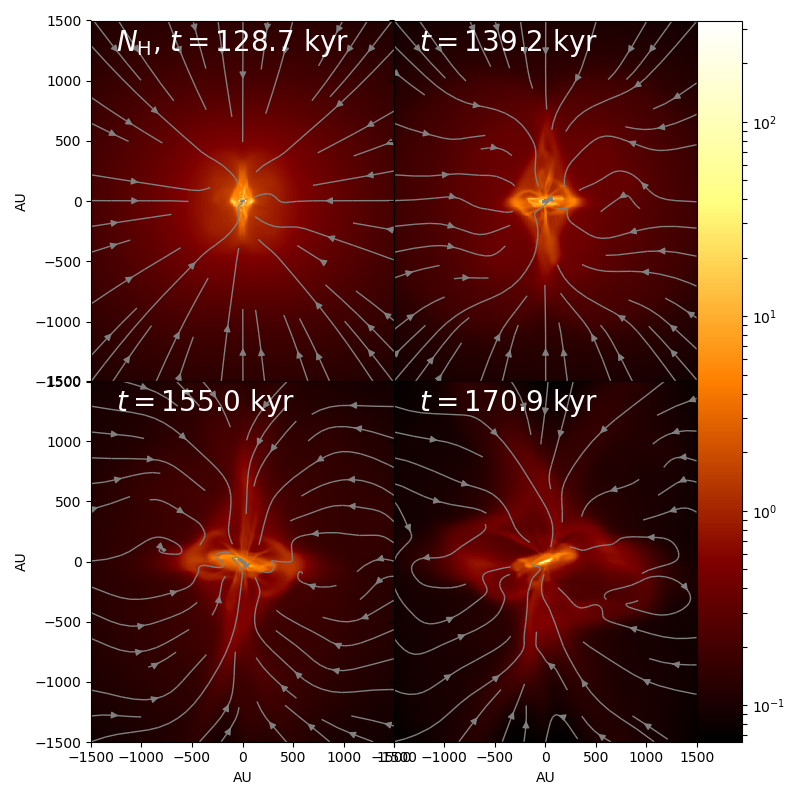}
\caption{Time development of the \textit{infunnel} of Model H in large-scale. The colors refer to column densities. The infunnel, an axial funnel-shaped infall stream, can be perceived as the cone-like enhanced column density extending in vertical direction. \label{fig:timeframe2}}
\end{figure*}

An \textit{LS} object (Model H) has very similar characteristics to the \textit{CS} case, but the system is more irregular, with smaller and less dense disk. Of particular note is, that the rotational plane of the system is tilted due to precession and a cone-like funnel-shaped axial infall structure is witnessed, which we call for short an \textit{axial infunnel}, coining a natural and useful term if it does not already exist. While the bulk of the accretion is concentrated on the disk plane, the effects of axial inflow are strong enough to be noticeable. A funnel-shaped infall can also be found in the CS case, when velocities are examined, but it has no clearly visible correspondence in the column densities.

The infunnel witnessed in the \textit{LS} case is irregular and shows variation of shape and size over time, making it especially apparent in movement (See Figure \ref{fig:timeframe2}). Looking at small scales, some precession along with the disk is noticeable (See Figures  \ref{fig:timeframe_precess} and \ref{fig:timeframe2}).  The infunnel can look deceptively like a conical outflow structure due to its visual shape and alignment with the disk; however, when actual flow lines are displayed, the direction of the flow towards the central core is clear. The outflow in \textit{LS} case is only sporadic and non-continuous. However, outflow ejection does open up the infunnel, likely due to the release of magnetic energy. This axial infunnel is further discussed in Section \ref{sec:infunnel}.

The other cases, however, do not form a spiral-like end state, or a disk. \textit{LA} objects have DEMS-like empty areas with loop-like edges extending outwards along the rotational axis from the central point. Viewed face-on, an arched bar-like pseudodisk is visible, with a X-shaped ridge around the central object, lacking a disk. \textit{LP} cases have similar effects, where the growing magnetic instability leads into development of DEMS with their loop-like edges. In motion, the DEMS circulate around to rotational plane of the system with one edge facing the central object. Increased initial magnetic field makes these edges of DEMS more stable and thick and the DEMS appear as even more visible holes in the density structure. \textit{LA} and \textit{LP} cases display wide continuous outflows, with \textit{LA} along the rotation axis, and \textit{LP} in the rotational plane perpendicularly to the initial magnetic field direction. Both are connected to the excessive magnetic braking. The ambiguous case with $45\degree$ misalignment, Model D, shows \textit{LP}-like outflow behavior.

In \textit{LP} and \textit{LA} cases there is no disk formed in the end. Due to the lack of disk formation, most of the further analysis in this paper will focus to \textit{CS} and \textit{LS} cases, for which most of the more detailed analysis is more functional and appropriate within the scope of this study.  

\subsection{Small scales ($R \sim 100$\,AU)}\label{sec:small}

\begin{figure*}
\plotone{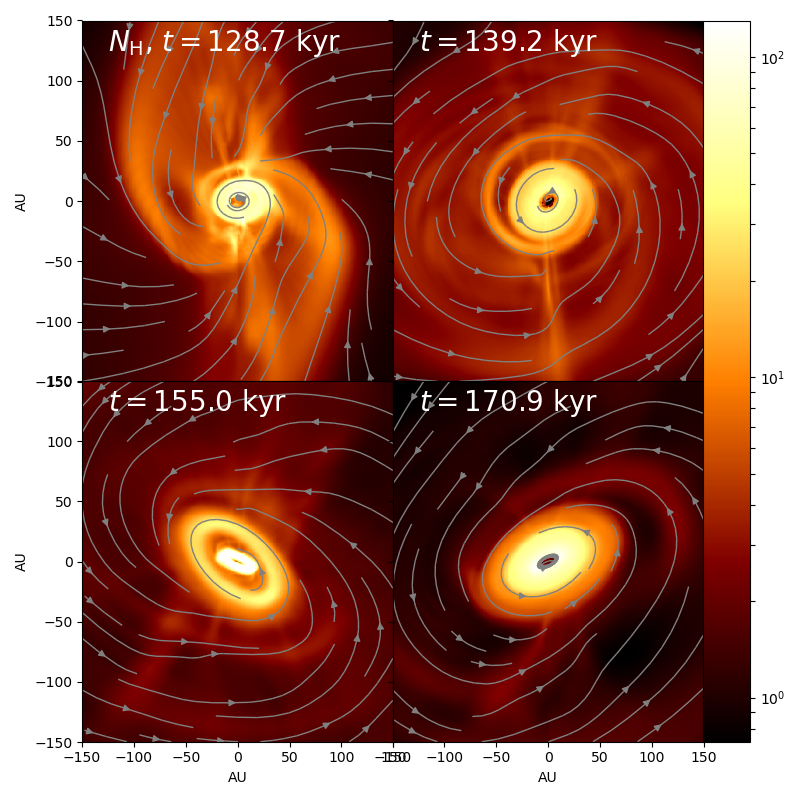}
\caption{Precession of the disk in Model H seen in small-scale at $45\degree$ angle. The colors refer to column densities. \label{fig:timeframe_precess}}
\end{figure*}

\begin{figure*}
\plotone{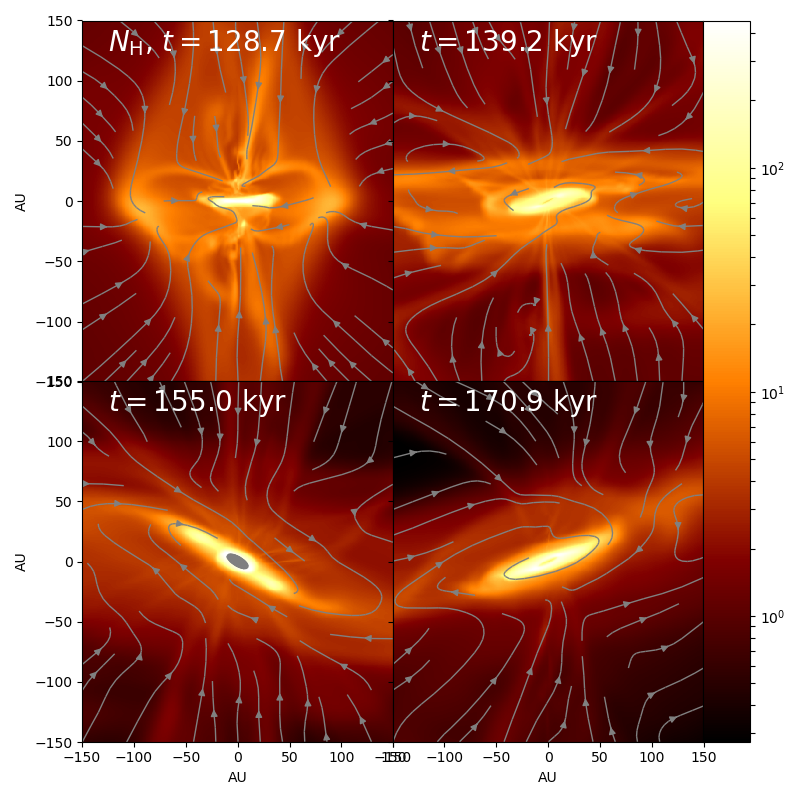}
\caption{Time development of the infunnel of Model H in small-scale. The colors refer to column densities. \label{fig:timeframe1}}
\end{figure*}

What makes the small scales noteworthy is the formation of a disk-like object at the center of the system in \textit{CS} and \textit{LS} cases (See Figures \ref{fig:close_above}, \ref{fig:close_45} and \ref{fig:close_edgeon}). The disk in \textit{CS} case is approximately 200\,AU in diameter and retains a spiral structure. In contrast, the 100\,AU diameter disk in \textit{LS} case is precessing around the rotation axis, which is the likely reason why its existence was not properly detected by LKS13, which relied on cutout slices from the model frames. A closeup of \textit{LS} precession is demonstrated in Figure \ref{fig:timeframe_precess}. In the \textit{CS} case, the disk has no visible precession during its whole development (see also Section \ref{sec:ts}), and the inner spirals of the disk wrap themselves around tightly.

The \textit{LS} case displays additional curious behavior: the disk often contains two differently precessing components --- best noticed when the object is observed in movement. The outer ring forms a band, which then precedes around the central protostar. The inner ring is denser than the outer ring and it precedes around the center with a slightly faster frequency than the outer ring. The third panel of Figure \ref{fig:timeframe_precess} demonstrates that occasionally a gap appears between the two rings. The inner and outer components are visible in Models H and P, but not in Z. 

In short, the CS case displays a larger disk 200\,AU  in diameter without visible precession while the LS case has a smaller 100\,AU disk, which shows steady precession over time. There is a noticeable jump in column density between the disks and their surroundings, and it is therefore reasonable to consider the disks to be distinct objects embedded within the rest of the system. This jump has been clearly detected in, e.g., HH212 \citep{Lee2017SciA}. It strengthens the case  for using high-resolution dust continuum imaging to detect disks in protostellar systems. 

\subsection{Polarization characteristics} \label{sec:diskpol}

Polarization calculated with \pers\ is directly connected to the magnetic field. Looking at Figures \ref{fig:close_above}, \ref{fig:close_45} and \ref{fig:close_edgeon}, the contrast between the disks and their surroundings in polarization is strong. This was to be expected, because both \textit{CS} and \textit{LS} disks wrap a strong azimuthal magnetic field around themselves. Therefore, Stokes $Q$ and $U$ appear regular in the case of \textit{CS} and \textit{LS}, where most of polarized emission is focused on the dense disk. 

In \textit{LP} and \textit{LA} cases irregularities are visible surrounding the DEMS; because magnetic pressure is strong, and a large part of the angular momentum is lost, the field no longer strictly follows a geometry driven by rotation. However, additionally noteworthy is the hourglass pattern seen in \textit{LA} case from above (Figure \ref{fig:all_above}), likely signifying the rigid behavior of the magnetic field in the large scales.    
In \textit{CS} and \textit{LS} systems, viewed from above (Figure \ref{fig:all_45}), the spiral structure can be traced corresponding to where the polarization degree is weak. Also, the polarization derived \textit{B-vectors} show hints of spiral geometry. The inner disk inherits the spiral-like nature of the system, but that is only weakly noticeable in polarization (Figure \ref{fig:close_45}). These features might not remain noticeable however, because spiral features are mostly visible in connection to the inflow envelope around the disk where, as will discussed in Section \ref{sec:reslimit}, the effect will be muddled by noise.

The polarization behavior of \textit{LS} resembles that of \textit{CS}. However, the geometry is more irregular with the disk being smaller and therefore having smaller surface area where the polarization is coherent. The rotationally supported disk shows a new feature: that polarization is connected to the precession of the disk. The precession is reflected in the polarization with B-vectors being aligned in the disk plane (Figure \ref{fig:close_edgeon}). 

In both \textit{CS} and \textit{LS} cases B-vectors show clear toroidal field direction when observed edge-on, in that they present similar kind of orientation as a typical edge-on disk \citep[See e.g.][for a recent dust polarization observation of HH212]{Lee2018}, or the surrounding pseudodisk. Therefore, observation of B-vectors will not differentiate the misaligned disk formation from more ordinary scenarios.  

In \textit{LP} polarization shows rotational characteristics, but growing in irregularity towards the center. When DEMS form, polarization follows the edges of the DEMS, tracing their borders. The DEMS can only be perceived from above, with B-vectors following azimuthal direction. It is noteworthy that the LP case shows B-vectors with apparently poloidal alignment in the small-scale (Figure \ref{fig:close_edgeon}), while the outflow displays toroidal B-vectors  (Figure \ref{fig:all_edgeon}). With \textit{LA}, near the vertical DEMS edge-like loops there is the strongest variation in polarization parameters. Seen from above, the \textit{LA} has B-vectors perpendicular to the bar-like density structure.

\subsection{Effects of instrument resolution}\label{sec:reslimit}

Having characterized the properties of our object types, the question remains, how limited resolution and signal-to noise ratio (SNR) affect the results. Many features seen by \pers\ can be muddled by telescope beam and sensitivity, therefore consideration for instrument limits is called for. Let us assume that the object would be relatively close by, at a distance of 120 pc and that the resolution corresponds to $0.25\arcsec$ \citep[as in the observation of VLA 1623A by][]{Sadavoy2018}. 

Let us then divide this into 4 levels of SNR 10, 100, 500 and 1000. In this cased SNR is determined as for Stokes $I$, $Q$, $U$ as 
\begin{equation}
    \mathrm{SNR} = \frac{\mathrm{max}(I_\mathrm{CS})}{\sigma_I}
\end{equation}
\begin{equation}
    \mathrm{SNR} = \frac{\mathrm{max}(|Q_\mathrm{CS}|,|U_\mathrm{CS}|)}{\sigma_{QU}}
\end{equation}
The fluctuations are scaled with the values of \textit{CS} (Model G), so that the noise level is similar in every case. In addition, both $Q$ and $U$ share the same level of noise. The noise itself is assumed to follow a Gaussian distribution. 
In addition, based on the determined $\sigma_{QU}$ we also calculate a basic maximum likelihood debiased polarized intensity\citep{Simmons1985, Vaillancourt2006}:
\begin{equation}
    P_I = \sqrt{Q^2 + U^2 - \sigma_{QU}^2}, 
\end{equation}
and we mask out values with $P_I < 4\sigma_{QU}$ as was done by \citet{Sadavoy2018}. 

Some object types are clearly more visible in regards to instrument sensitivity (see Table \ref{tbl:SNR}). A \textit{CS} object is the most coherent and visible. \textit{LA} and \textit{LP} cases require the highest sensitivity. In principle this would mean that disk and some of the surrounding inflow formed in the misaligned magnetic field situation should be resolved, even some of the spiral characteristics should be visible. However, if the system is closer to \textit{LS} case, all the features become more difficult to recognize, because the \textit{LS} disk is less dense and coherent, with smaller diameter.

DEMS are not likely observable, as they require high sensitivity to be seen. Considering also the loss of extended emission in interferometric observations, it would be highly unlikely that the edges of DEMS could be distinguished from other kinds of filament-like irregularities, lacking sufficient contrast.

\begin{table}
\begin{center}
\caption{Visible features depending on the SNR. CO = Compact object, Disk-like object = DLO, SS = spiral structure, Polarized structure = PS, DI = disk inclination.\label{tbl:SNR}}
\begin{tabular}{rllll}
\tableline
\tableline
SNR  & CS & LS & LA & DL \\  
\tableline
8    & CO, PS & CO & --- & --- \\  
16   & CO, PS & CO & --- & --- \\  
32   & DLO, PS & CO, PS & --- & --- \\  
64   & DLO, SS, PS & CO, PS & --- & --- \\  
128  & DLO, SS, PS & DLO, DI, PS & --- & --- \\  
256  & DLO, SS, PS & DLO, SS, DI, PS & Bar, DEMS & DEMS \\  
512  & DLO, SS, PS & DLO, SS, DI, PS & Bar, DEMS & DEMS \\  
1024 & DLO, SS, PS & DLO, SS, DI, PS & Bar, DEMS & DEMS \\
\tableline
\end{tabular}

\end{center}
\end{table}

\begin{figure*}
\plotone{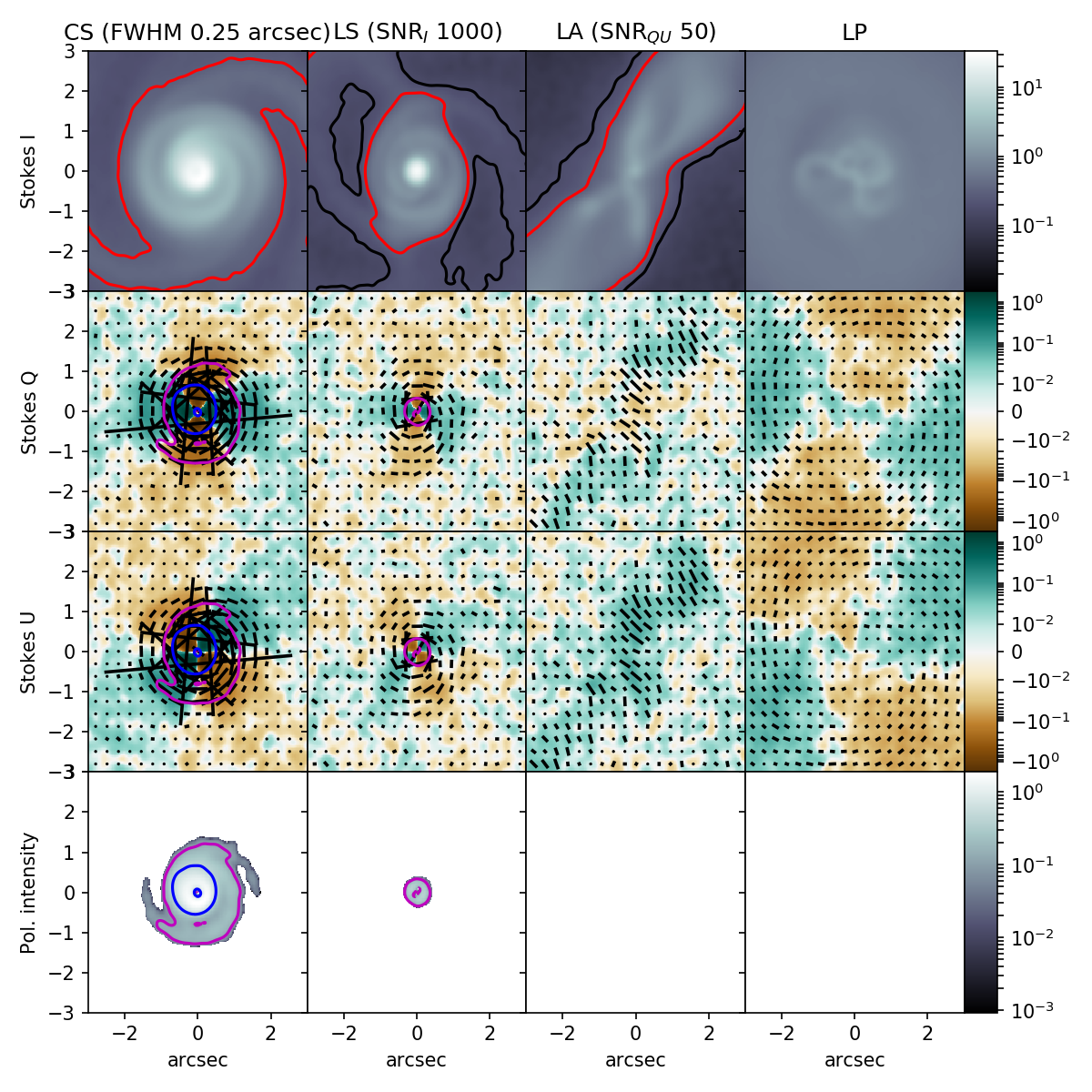}
\caption{Models G, H, I and A viewed from above, 
with 0.25 arcsec FWHM and SNR $\sim 1000$ and SNR $\sim 50$ for Stokes $I$ and $Q/U$ respectively. The red contour shows the regions with $5\sigma$ detection and black with $3\sigma$. The magenta and blue contour shows where $\mathcal{P}_I/\sigma_{\mathcal{P}I} \geq 1$ and $ 4 $ respectively. \label{fig:noise128_0}}
\end{figure*}

\begin{figure*}
\plotone{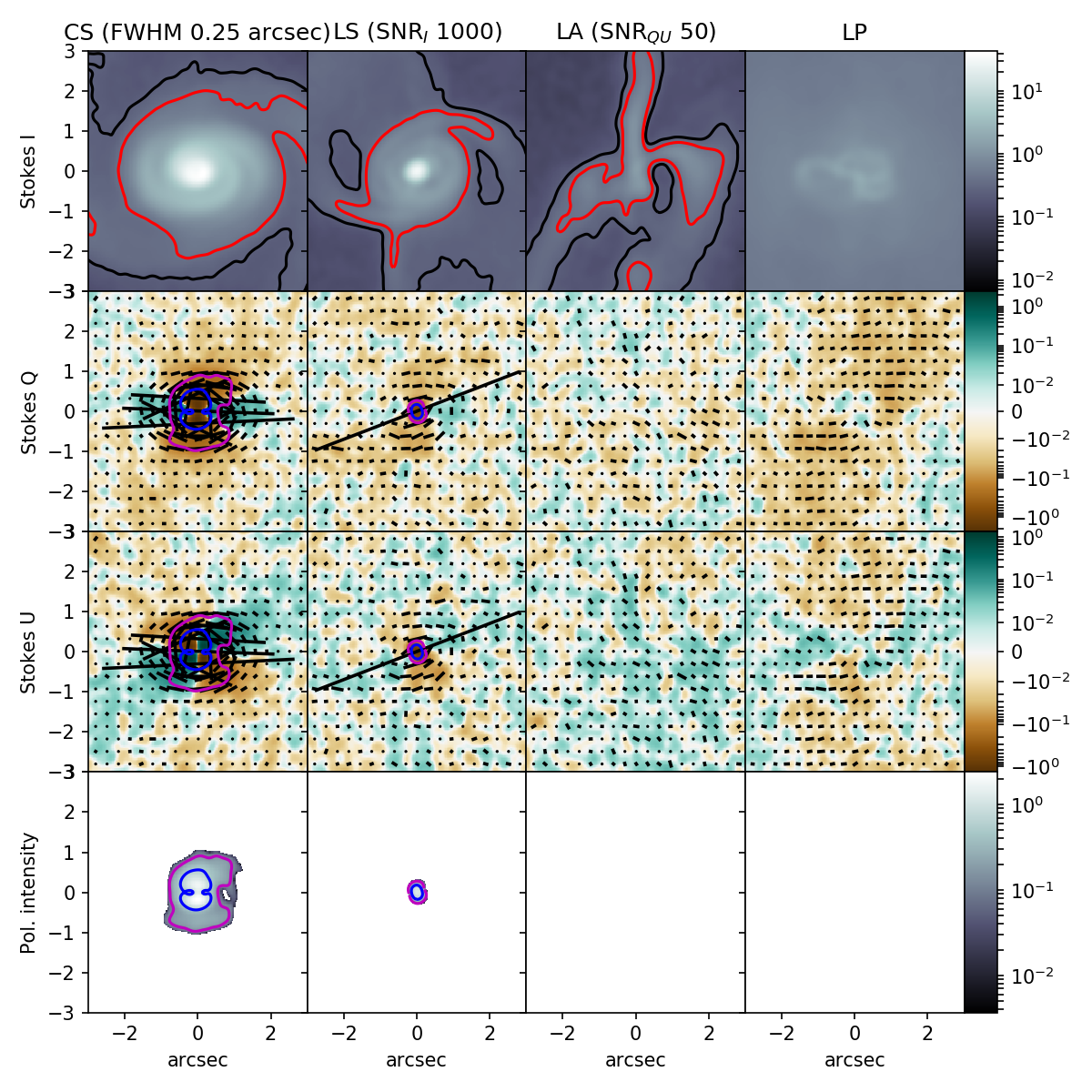}
\caption{Models G, H, I and A viewed from 45 degree inclination, with 0.25 arcsec FWHM and SNR $\sim 1000$ and SNR $\sim 50$ for Stokes $I$ and $Q/U$ respectively. \label{fig:noise128_45}}
\end{figure*}

\begin{figure*}
\plotone{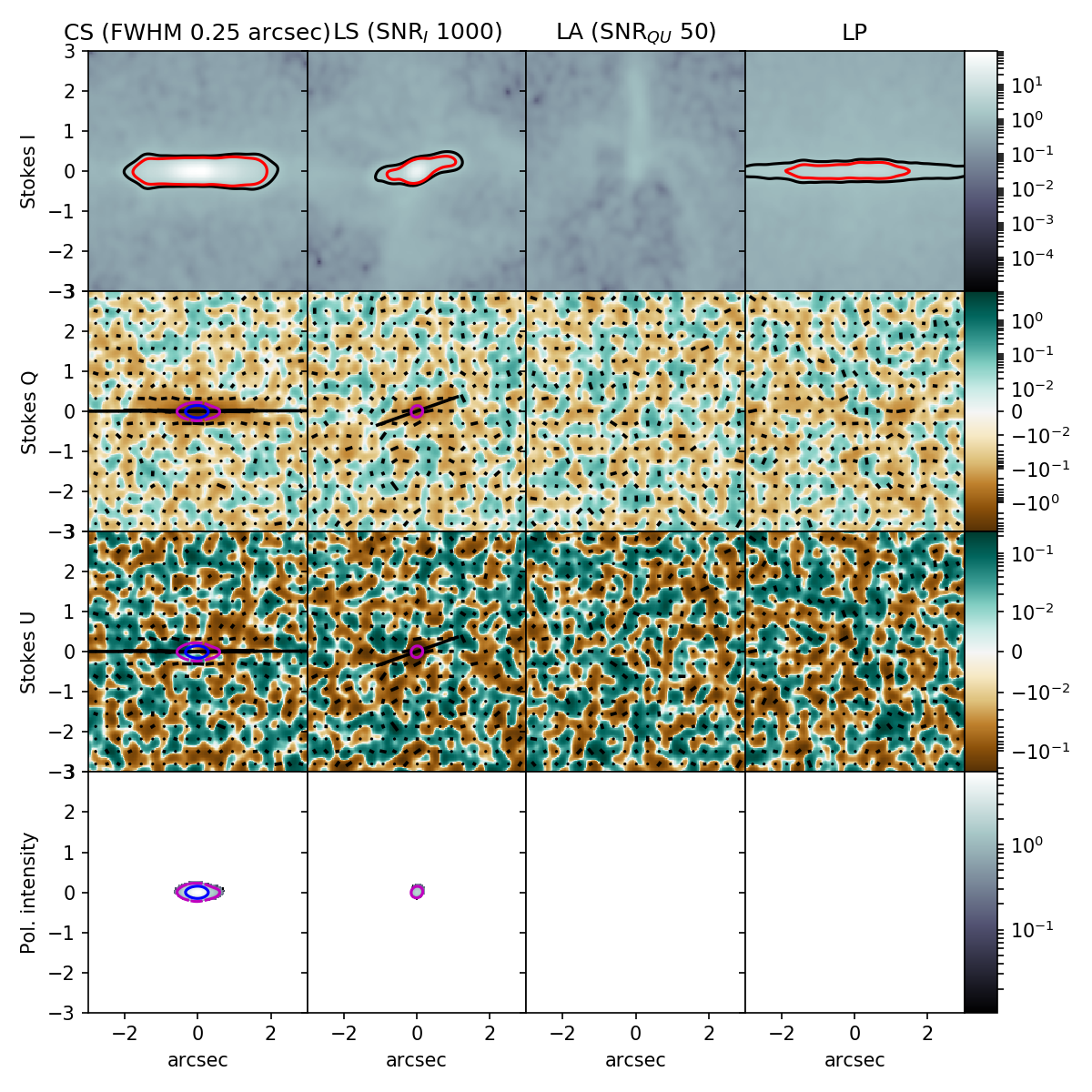}
\caption{Models G, H, I and A viewed from 45 degree inclination edge-on, with 0.25 arcsec FWHM and SNR $\sim 1000$ and SNR $\sim 50$ for Stokes $I$ and $Q/U$ respectively. \label{fig:noise128_90}}
\end{figure*}

In more specific case, to approximate the sensitivity of \citet{Sadavoy2018}, we get SNR $\sim$ 1000 for $I$ and SNR $\sim 50$ for $Q$ and $U$. Results are shown in Figures \ref{fig:noise128_0}, \ref{fig:noise128_45} and \ref{fig:noise128_90}. It seems, therefore, that the corresponding ALMA resolution and sensitivity should be in principle enough to recognize traces of the spiraling gas, not counting complexities caused by interferometry. However, polarization itself would only give an impression of a simple toroidal magnetic field, with noise making any spiral geometry practically invisible. 

\subsection{Time development: Gaussian Analysis}\label{sec:ts}

Following the methods described in Section \ref{sec:gauss}, we get some results describing the characteristic behavior of the system. We have left out the \textit{LA} and \textit{LP}-type systems from this analysis. Those do no produce a disk --- making the Gaussian fitting method unreasonable. 

\begin{figure}
\plotone{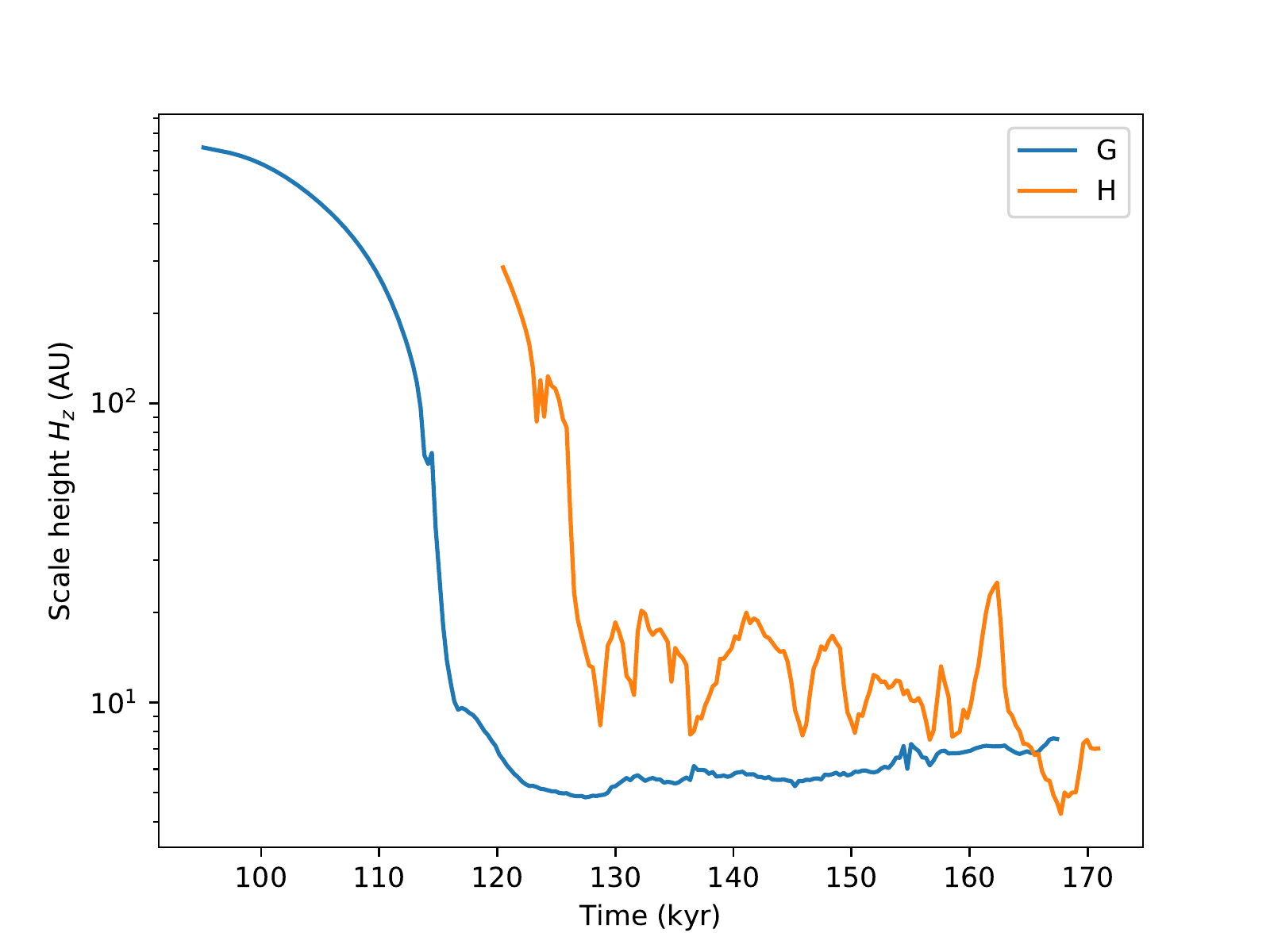}
\plotone{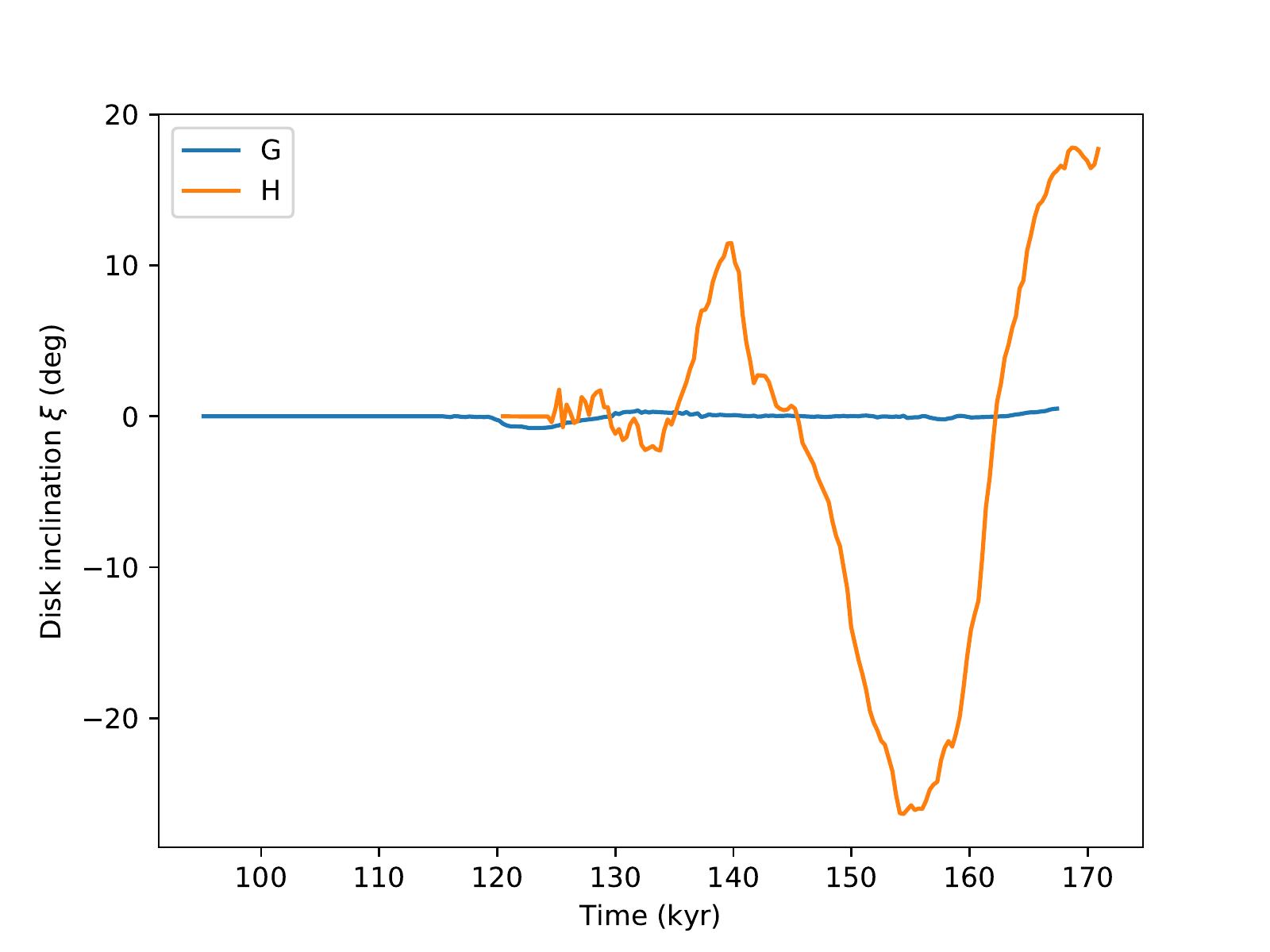}
\plotone{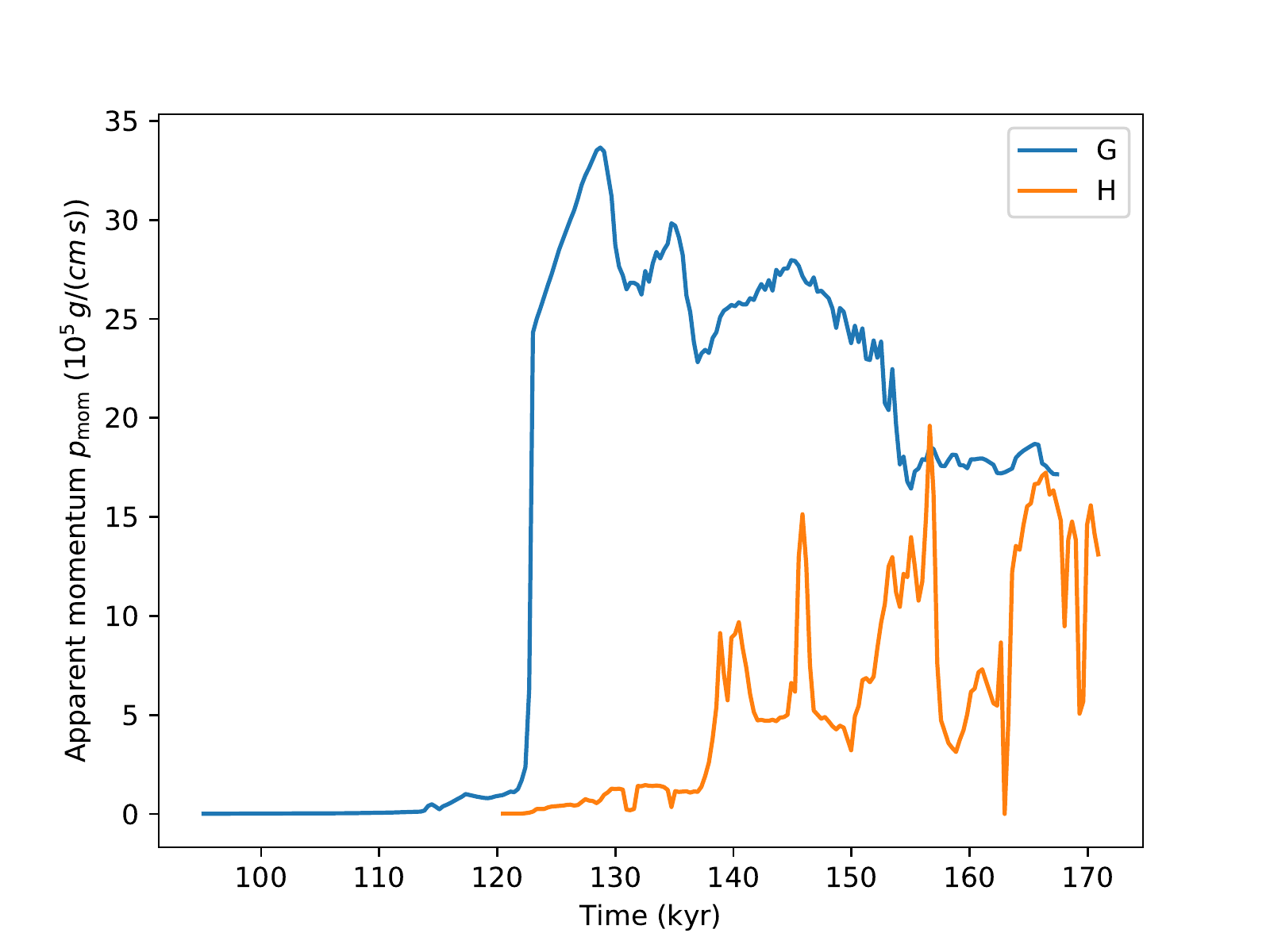}
\caption{Evolution of scale height $H_{z}$ (\textit{Top}), precession angle $\xi$ (\textit{Middle}) and apparent momentum $p_\mathrm{mom}$  (\textit{Bottom}) as a function of time.
\label{fig:timeseries}}
\end{figure}

In all cases, after a visible disk-like object has emerged from the initial envelope, essentially when there is enough mass in the central protostar to be gravitationally significant, the scale height remains around an approximate average value, and these averages range between $H_{z} \sim 5$ to $\sim 20$ AU for Model H and at $H_{z} \sim 6$ AU for Model G (See Figure \ref{fig:timeseries}, Top). The higher variability of Model H is most likely due to its precession, which changes its apparent width somewhat in regards to the fitting method. As will be discussed in Section \ref{sec:spiralflow}, magnetic field might be a reason for such a stable scale height. 

Model G does not show any kind of tilting, the disk almost perfectly aligned with the horizontal direction for the duration of the whole simulations (See Figure \ref{fig:timeseries}, Middle). However, Model H shows clear precession, with a precession angle of $\xi \sim 20\degree$ when the system is most coherent towards the end, and the timescale of a single round of precession is $\sim 30\kyr$. When observed in motion, the precession is obvious as soon as a resemblance of a disk is seen. 

\begin{figure}
\plotone{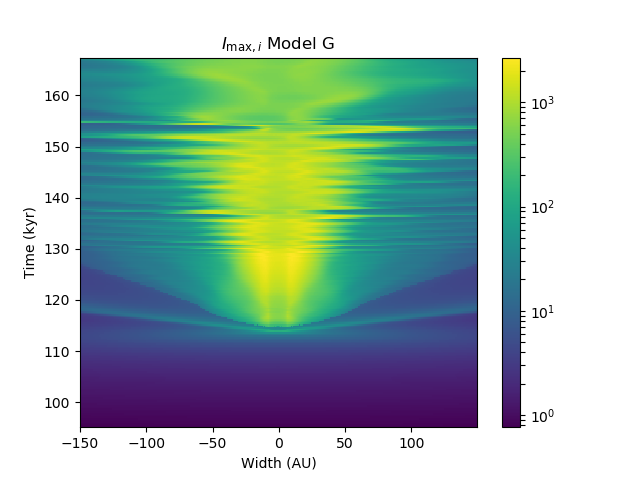}
\caption{$I_{\mathrm{max,}i}$ as a function of time for the Model G . \label{fig:GmaxI}}
\end{figure}

In Figure \ref{fig:timeseries} (bottom) we display the change of apparent momentum $p_\mathrm{mom}$ over time. In Model G, after initial growth the momentum decreases, with the strong changes reflecting the moments when mass is absorbed by the central object, therefore disappearing from the disk itself. The irregularities in Model H might be partially due to the projection effects caused by precession. However, in general the Model G object has more momentum, reflecting the stronger magnetic braking in Model H (LKS13).    

\begin{figure}
\plotone{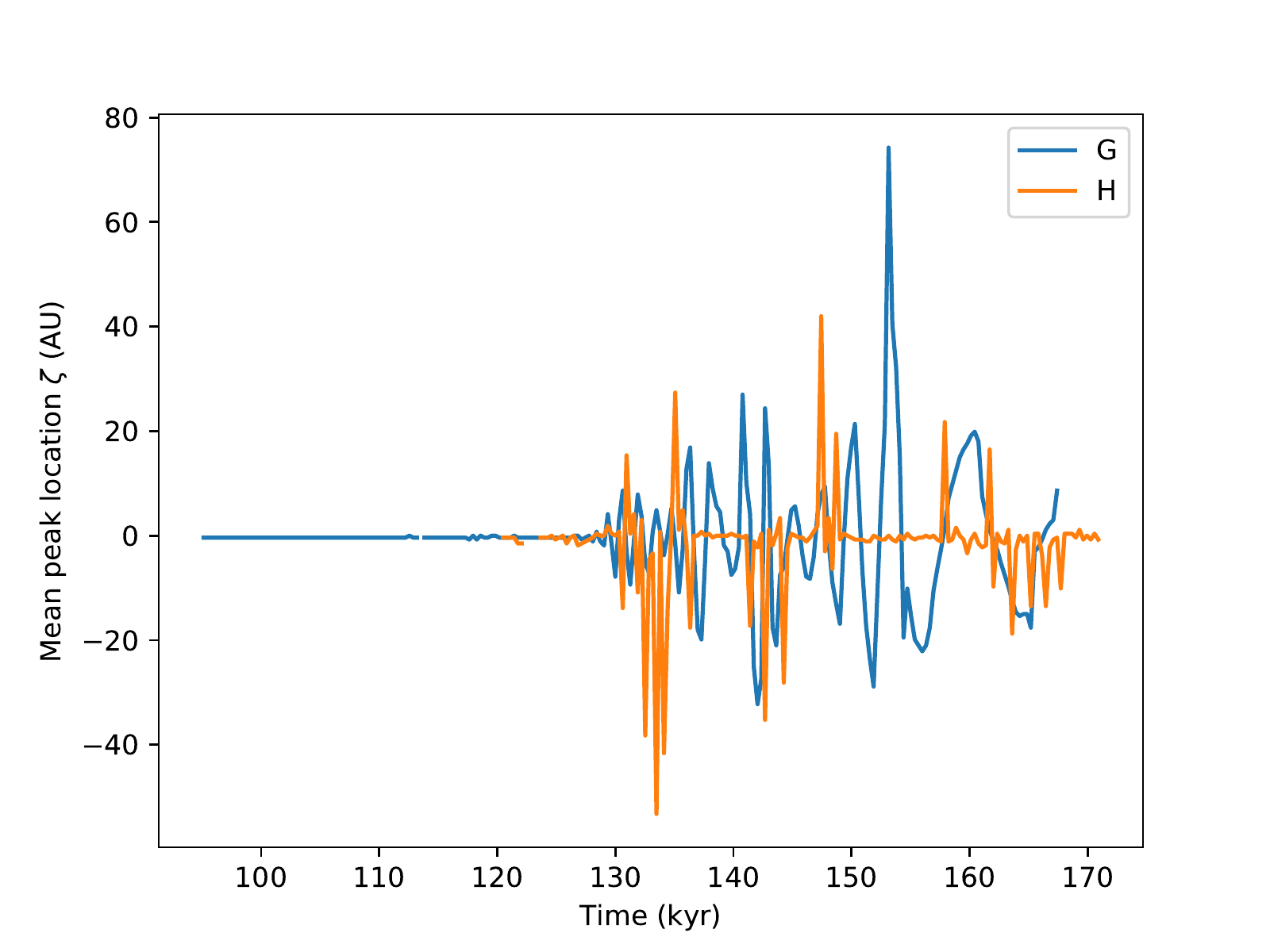}
\plotone{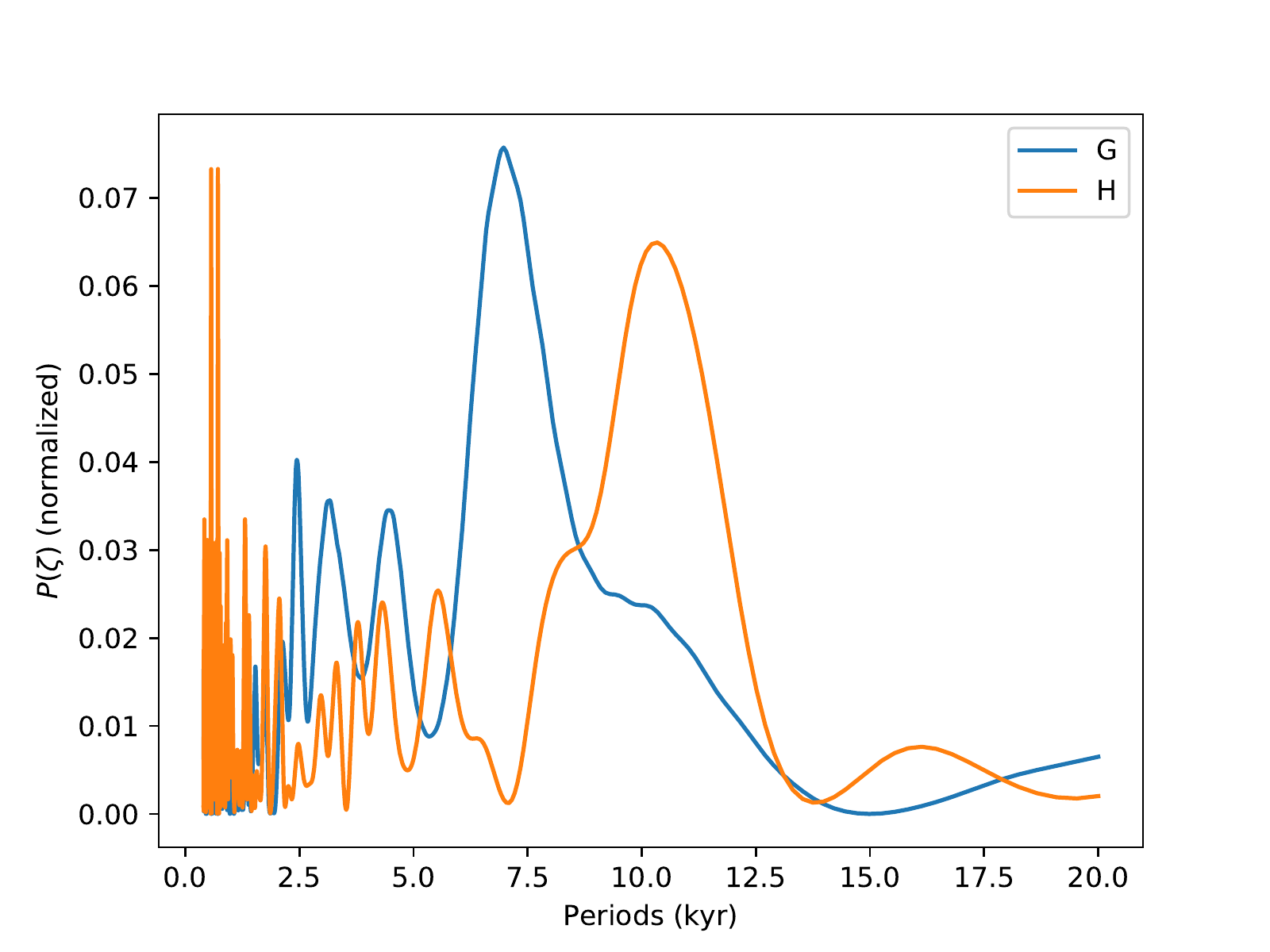}
\caption{Mean peak location $\zeta$ as a function of time (\textit{Top}). Normalized Lomb-Scargle periodograms $P(\zeta)$, as a function of period $T$ (\textit{Bottom}).  
\label{fig:lombscargle}}
\end{figure}

Figure \ref{fig:GmaxI} displays fitted $I_{\mathrm{max,}i}$ as a function of time for the Model G, as an example. These can be reflective on the oscillation of spirals, specifically their projection. In the Figure \ref{fig:lombscargle} (top), the mean peak location $\zeta$ is traced, showing for Model G a pattern of oscillations, not seen in the case of Model H. This is likely a combination of effects from both wrapping up of magnetic field lines and self-gravity.  To make this more clear, we took a standard Lomb-Scargle periodogram of $\zeta$, getting Figure \ref{fig:lombscargle} (bottom). A Lomb-Scargle periodogram is a common method of calculating power-spectra of an unevenly spaced time-series within a set range of periods \citep{Lomb1976, Scargle1982}.
Model G shows a peak at $T \sim 7.5$ kyr, with a smaller peak between $T \sim 2.5$ and $5$ kyr. As such, this would support the visual observation that the spiraling disk several shorter and longer oscillatory modes as discussed in the Section \ref{sec:large}. Model H shows comparable behavior with the main peak at $T \sim 10$ kyr and smaller peak found in similar range as for Model G. However, $\zeta$ is less clearly resolved. 

\subsection{Velocity}\label{sec:velocity}

LKS13 found the Model G to produce a rotationally supported disk and we would add Model H into the list, both corresponding what we call here CS and LS type objects. While \pers\ is not capable of genuinely simulating radial velocities measured from the spectral lines, we can still explore how flows within the core relate to what is visualized. 

\begin{figure}
\plotone{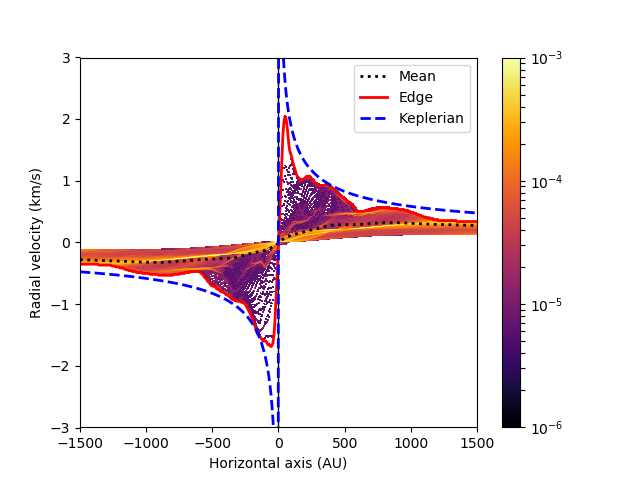}
\plotone{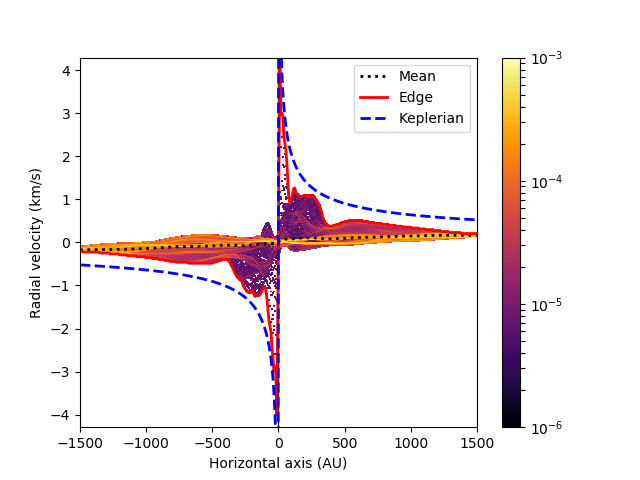}
\caption{PV-diagrams of Model G (top) and Model H (bottom). The colors represent probability density with the total pixel count. The red line denotes largest velocities at the given horizontal coordinate, and the hatched line traces $\vkep$ based on $M_*$. 
\label{fig:pvdiagram}}
\end{figure}

As described in Section \ref{sec:meanvel}, 
we calculated the density weighted line-of-sight averages of all Cartesian velocity components. As radial velocities $v_\mathrm{LOS}$ are what in principle could be directly observed, we constructed a position-velocity (PV) diagram for the edge-on models using the horizontal axis equivalent to the numerical midplane. Figure \ref{fig:pvdiagram} shows the PV-diagrams of Models H and G, and in both cases their rotational profile is fairly similar. 

The velocity profile with a Keplerian power law index,
\begin{equation}
    \vkep = \sqrt{\frac{2GM_*}{R}} \label{eq:keplervel}
\end{equation}
is calculated based on the mass of the central point ($M_*$). As can be seen, the fastest velocities in the distribution, which trace best the rotating disk velocities, approach Keplerian, where the curve behaves approximately, but not absolutely, as an upper limit. Therefore, the disks are not strictly Keplerian, in respect to the central point mass, with the spiral structures in particular leading to systematic deviation from circular Keplerian velocity, discussed more below. 

\begin{figure}
\plotone{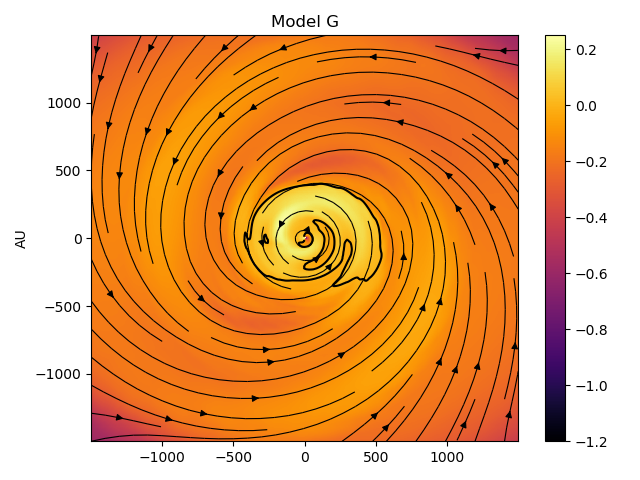}
\plotone{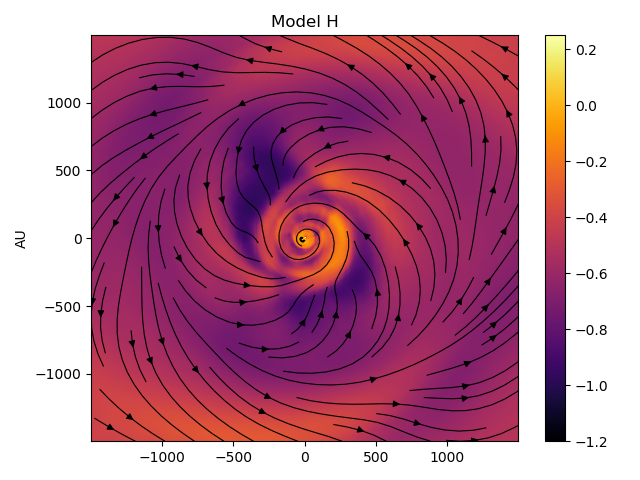}
\plotone{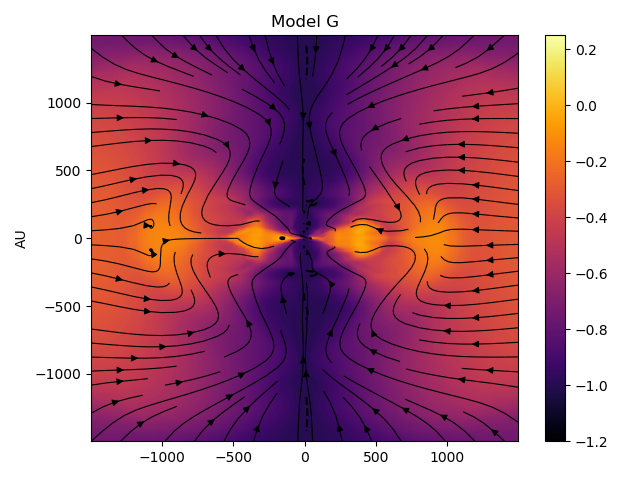}
\plotone{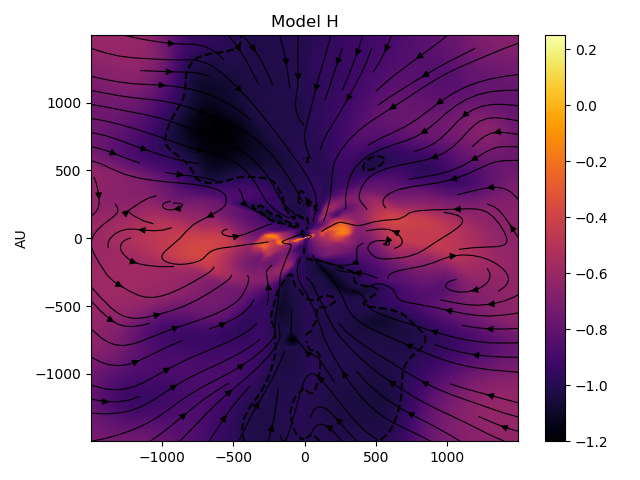}
\caption{$\delta \vrot$ of Model G (First) and H (Second), viewed from above; and $\delta \vrot$ of Model G (Third) and H (Fourth), viewed edge-on. It should be noted that in the edge-on case $\vkep$ is calculated relative to the origin of the horizontal axis instead of the central point. 
\label{fig:deltavdisk}}
\end{figure}

To explore non-Keplerian disk behavior, we analyzed the relative deviation from Keplerian power law, based on the analysis of \citet{Teague2018}. Therefore we calculate $\delta \vrot$ with 
\begin{equation}
    \delta \vrot = \frac{\vrot - \vkep}{\vkep} 
\end{equation}
based on the line-of-sight velocity averages and assuming that the system rotates around the central point mass in counterclockwise direction. $\vkep$ is cylindrically symmetric, lacking eccentricity in its orbital direction. We denote $\vrot < \vkep$ as `sub-Keplerian' and $\vrot > \vkep$ as `super-Keplerian' motions. The result for Models G and H are shown in Figure \ref{fig:deltavdisk}. We can see that only the spiral ridges in the velocity patterns have rotational velocities approaching the corresponding circular Keplerian velocity and at some points --- like parts of the inner disk of the Model G --- $\vrot$ can be even faster than $\vkep$ itself (leading to the transient orbiting behavior discussed in \ref{sec:large}). Therefore, magnetic forces and self-gravity contribute to the rotational velocities of the model disk in significant ways both enhancing and restricting it. Seen edge on, we see torus-like quasi-Keplerian regions which correspond to the visible projection of the spirals. 

\begin{figure}
\plotone{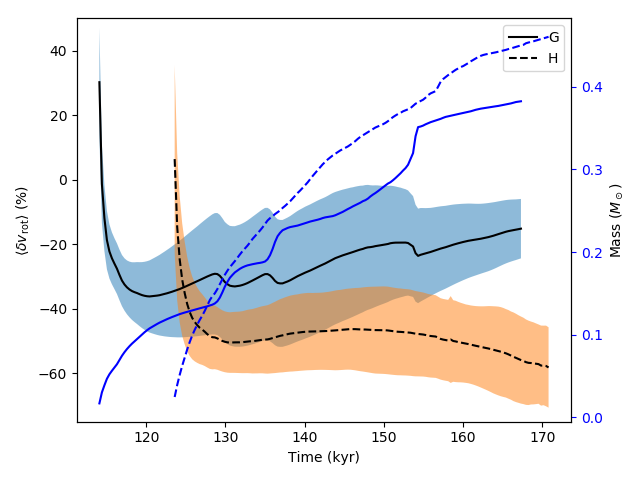}
\caption{Time development of $\langle \delta \vrot \rangle$, starting from the point where $M_* > 0.01 M_\odot$. The colored areas show the corresponding range of one standard deviation, $\sigma_{v,\mathrm{rot}}$.  
\label{fig:deltavrms}}
\end{figure}

The development of $\delta \vrot$ in time is described in Figure \ref{fig:deltavrms}, with the rms-values ($\langle \delta \vrot \rangle$) calculated from $\delta \vrot$ maps, viewed from the top like in Figure \ref{fig:deltavdisk}. The value of $\langle \delta \vrot \rangle$ does not change considerably from the beginning. The quick variation of $\langle \delta \vrot \rangle$ in Model G corresponds to the occasionally rapid mass accumulation by the central object. Considering also the difference in the magnetic field strengths, the stronger field leads to larger difference from circular Keplerian velocities in respect to the central mass.

In Figure \ref{fig:deltavdisk}, we also show the plane-of-the-sky streamlines for the flows in the system. While these cannot be directly observed without sufficient inclination based on radial velocities, they reveal a notable pattern in the inflow in relation to rotation. First, due to magnetic forces around 1000 AU radius, the horizontal inflow drops in $\sim 90\degree$ angle towards the midplane. Second, the flow spirals downwards towards the central mass along the horizontal plane, forming a pattern of accretion. These features can be best understood by looking into the phenomenon of pseudodisk, which will be further examined in Section \ref{sec:spiralflow}.

\section{Discussion}\label{sec:disc}

We have categorized the numerical results of LKS13 into four types based on 3D time evolution with \pers. While we give consideration to all of them for their own right, it is apparent that only \textit{CS} and \textit{LS} models show signs of disk formation and therefore will be discussed in more details for their observable features. While individual dynamical evolution carries a strong dependence on the initial conditions, polarization effect alone may not secure a conclusive distinction. Velocity profiles with deviations from a circular Keplerian velocities can tell about the effects of magnetic braking done to the systems.
Recent results from Class 0 objects VLA 1623A and IRAS 16293 can also serve as convenient references for comparisons.

\subsection{Hydromagnetic spiral inflow}\label{sec:spiralflow}

The \textit{CS} and \textit{LS} disks and their surroundings are systems of magnetically confined flows. 
The general pattern of rotational inflow (See Section \ref{sec:velocity} and Figure \ref{fig:deltavdisk}) flows inwards in a curved trajectory with two arms directed by the analogously structured magnetic field. 
This phenomenon is known as the pseudodisk which LKS13 described as a ``snail shell'' shape, also denoting the pseudodisk as ``pseudospirals''. Coined by \citet{GalliShu1993}, a pseudodisk is a disk-like inflow structure which results from magnetic inflow following pinched field lines of the collapse. 
A pseudodisk superficially resembles the disk, but it is not fully rotationally supported. Rather its motions are dominantly supersonic inflows, typically slowed down from free-fall speeds by a combination of rotation and magnetic tension forces. 
However, the general direction of the field lines in our \textit{CS} and \textit{LS} cases is orthogonal to the \citet{GalliShu1993} scenario, perhaps pronouncedly so as \citet{GalliShu1993} being pioneering work did not include (misaligned) rotation. 

In the case of \textit{CS}, the rotating vertical inflow is strongly streaming towards the pseudospiral ridges already at $R \sim 1000$\,AU.
In the \textit{LS}, the effect is there but it is more chaotic. These would demonstrate that the magnetic forces are directing the collapse dynamics in a manner of a pseudodisk, after the initial magnetic field has sufficiently condensed. This could be the reason why the approximate disk height is determined already early on in the development (Figure \ref{fig:timeseries}), with the magnetic field directing the system into such a configuration. 
A noteworthy feature is that when the inflow streams along the pseudospirals that connect to the spiraling inner disk, there is no clear presence of a centrifugal shock. The transition between disk and pseudodisk is smooth, in contrast to the more standard pseudodisk scenario with aligned rotation and magnetic fields \citep{KK2002}. 

In both CS and LS cases is a trailing spiral inflow of matter which is visible in the column densities (Figures \ref{fig:all_above}, \ref{fig:all_45}, \ref{fig:close_above}, \ref{fig:close_45} and \ref{fig:timeframe_precess}), leading from the lobes of pseudospirals towards the inner disk, which is potentially observable.
LKS13 argue that the spiral is strongly influenced by magnetic torques, and the spiral inflow is an essential part of the collapse process, as this is how inflow is most efficient.  Indeed, similar pseudodisk behavior is even present in the \textit{LA} case, where there is a (less bent) pseudospiral but no disk. However, as the system extends from the inner spirals of the disk to the infall lobes of the pseudospirals, they becomes increasingly difficult to observe due to lack of density and therefore emission. Despite demands on sensitivity, detecting parts of the pseudodisk should be considered a priority for detecting the magnetically misaligned collapse, as it is one of the most distinctive features of the model. 
While the spiral structures are visible in polarization (See Figures \ref{fig:all_above}, \ref{fig:all_45}, \ref{fig:close_above} and \ref{fig:close_45}), the effect is weak and can be very difficult to detect due to the distorting influence of noise (See Figures \ref{fig:noise128_0} and \ref{fig:noise128_45}).
Because any rotating disk system can be expected to have wound up toroidal magnetic field due to rotation, merely observed toroidal field cannot conclude its origin. 

Most of the above features will be visible with reasonable inclination, but the features of spirals and pseudospirals will become less conclusive when viewed edge-on, with the inner part resembling merely a disk-like object. However, the pseudospirals are effectively visible as additional low-density lobes in the midplane, if they are favourably aligned with the LOS. Finding such features around the edge-on disk observation would support misaligned collapse scenario, although this would require more careful assessment of the object in question. 

\subsection{Hydromagnetic precession}

The \textit{LS}-case disk shows precessing motion (See Section \ref{sec:small} and Figure \ref{fig:timeframe_precess}). The magnetic field is likely the primary cause of this phenomenon, because the weaker field \textit{CS} case does not show precessing behavior. LKS13 found with Models P and Q that, what they called ``porous'' disk, or our \textit{LS} case, would emerge with approximately $\lambda > 4$. Unfortunately, they did not probe the transition from \textit{CS} case to H \textit{LS} case, and therefore we cannot know at which stage the dynamical precession behavior emerges after the magnetic field strength increases.

A possible explanation for the precession could be found based on the anisotropic nature of the magnetic effects.
Consider a case of wound-up magnetic field: from initial collapse, the magnetic field is dragged by the collapsing flow into the typical hour-glass geometry and field lines are wrapped up by the rotation. Over time this magnetic tension tries to relax itself, but because the angular momentum is strong, the energy is released into the vertical motions leading to precession-type oscillation. However, investigating this conclusively is beyond the scope of this study.

\subsection{Axial infall streams and outflows. The infunnel.}\label{sec:infunnel}

In Section \ref{sec:large} we observed that some matter flows in through the polar directions like a magnetic funnel-shaped infall stream (for short, an \textit{axial infunnel}) along which the infall material spirals towards the central object itself (See Figures \ref{fig:timeframe_precess}, \ref{fig:timeframe1} and \ref{fig:timeframe2}). The infunnel can lead to some confusion, if the flows themselves are not properly examined. Particularly, the shape of the infunnel in \textit{LS} can resemble an outflow cavity. If only the gradient of velocity is seen, without other kind of supporting evidence for an outflow, inflow could be mistaken for an outflow. This could be a point of concern when tracing outflow structures in smaller scales \citep[e.g.][]{Bjerkeli2016}, where observations are at the limits of instrument sensitivity. Even when observed in movement without proper velocity information, the edges of the infunnel appear seemingly like that of an outflow, because there are wave motions moving outwards from the center. However, the direction of these waves is opposite to their flow direction.

Outflows in the \textit{LS} case, on the other hand, only appear as bullet-like, sporadic pulses of ejected matter, which follow the inclination of the disk. A potential mechanism for the phenomenon could the magnetic tower jet \citep{LyndenBell2003, Kato2007}, as it primarily dependent on the wrapping of toroidal magnetic field. The mechanism could lead to the release of angular momentum after sufficient magnetic energy has gathered. LKS13 does not make note of the outflow in the Model H, but it is clearly visible from the time development of velocities and columns density. However, we cannot know how the outflow would behave over a long period of time.

\subsection{Patterns of rotation}

Another supportive signal for misaligned collapse would be found if some rotational velocities of the disk could be estimated. The wave like perturbations from a velocity profile with a Keplerian power law index in respect to the central mass, depicted in Figures \ref{fig:pvdiagram} and \ref{fig:deltavdisk}, could provide such signatures. Such deviations from circular Keplerian velocities will appear despite the accurate assessment of the central mass. However, the a fit to the velocities with a Keplerian power-law faces a number of challenges if the intent is to understand the kinematics of an observed object. In addition to the issues raised here, \cite{Harsono2015} have performed CO line modelling based on a part of the LKS13 dataset (Models A and G) giving some sense of observing the velocities in practice. 

Our PV-diagrams demonstrate one of the concerns. Most of the radial velocities are seen below the Keplerian power-law predicted by the mass of the central object, which in our case is well known. Physically this is reasonable: magnetic torques slow down rotation (LKS13) and there is substantial inflow towards the center. However, if a Keplerian power-law fit is done to an observed PV-distribution, where $M_*$ is not known, $M_*$ could be underestimated, because the disk is consistently sub-Keplerian on average as shown in the Figure \ref{fig:deltavrms}, with occasional components super-Keplerian with respect to the central mass. We also known that the rotation in the disk is not only influenced by the central object mass but by self-gravity of the disk. Therefore, development of kinematic spiral flow models, for PV-diagram fitting, could prove to be useful. 

Our spiral flows also demonstrate that circular Keplerian velocity assumption can be simplistic. The nature of the spirals in our collapse scenario is likely hydromagnetic, as described in Section \ref{sec:spiralflow}. Formation of spirals in a disk is in principle possible with purely hydrodynamical accretion streams during collapse \citep{Hennebelle2017}. Due to increase in kinetic energy, the inner disk is not likely directed merely by magnetic forces, but LKS13 point out, the snail shell shaped pseudodisk significantly directs the inflow of the collapse, making it at least partially responsible for the shape. In addition, a stronger magnetic field can inhibit rotation, as is seen in the \textit{LS} case. Therefore, the system as it is would not be the same without the magnetic fields. The pseudospirals surrounding the system present corresponding type of velocity perturbations as the spirals in the rotational supported disk, with dense parts approaching the circular Keplerian/super-Keplerian velocities with respect to central mass and low-column density areas appearing sub-Keplerian in \textit{CS} case.

The apparent variation of rotational velocities (between above and below the circular Keplerian speed $\vkep$) in the rotationally supported part of the disk can be best understood with eccentricity. The disk is approximately Keplerian with a component of accretion, but this rotation is not circular. Instead, the velocity flows follow eccentric orbits. When the eccentric orbits approach periastron, the fastest rotation velocities can be found, slowing down near apastron. They also correspond well with positive and negative radial velocities with respect to the center. Therefore, a significant part of the deviations are due to the fact that while rotation approaches the circular Keplerian velocities as defined in Equation \ref{eq:keplervel}, \textit{the rotation is not circular}.

\subsection{Planets?}

Spirals can be formed also by other processes than the magnetic collapse. Our velocity analysis was inspired by the methods used in studies of spiral structures that can emerge due to the influence of giant planet formation in the disk due to pressure gradient \citep{Kanagawa2015, Teague2018, Perez2018}. Using an equivalent definition of $\delta \vrot$ as in \citet{Teague2018} and \citet{Perez2018}, we can attempt to compare how the magnetically driven spirals differ from planet driven ones. 
However, it should be emphasized that the disks in our MHD models are of significantly earlier stage than even the youngest objects where planet formation is currently concerned, such as HL Tau \citep{Testi2015}. What this comparison does is to give an intuitive comparison to the magnitude of our spiral phenomenon. 

\citet{Teague2018} based their estimate on matching observed velocity profiles with a hydrodynamical disk model. Their fit of two Jupiter mass planets generate deviations from the circular Keplerian velocity, $\delta \vrot$, between $\sim -6 \%$ and $\sim 3 \%$. On the other hand, \citet{Perez2018} models their synthetic observations with planets as massive as $10$\,$M_\mathrm{Jupiter}$, generating $\sim \pm 20 \%$ velocity perturbations in $\delta \vrot$. 
While the perturbation in \citet{Teague2018} are relatively weak, the hydromagnetic spirals of our model can appear as perturbations of roughly equivalent magnitudes compared to the \citet{Perez2018} model, especially in the inner disk (See Figure \ref{fig:deltavdisk}). 
We cannot know, however, how well our velocity perturbations will last further as the disk develops to a stage where planet formation is possible. 

\subsection{Comparison to observations}

We have considered VLA 1623A \citep{Murillo2013, Sadavoy2018, Harris2018} and IRAS 16293 \citep{SadavoyIRAS2018} as potential candidates for misaligned formation. As more high-resolution observation of Class 0 objects are published, perhaps more potential candidates can emerge, but these two can function as an example of current possibilities and limitations. 

\citet{Sadavoy2018} observed VLA 1623A in ALMA Band 6 (1.3 mm) continuum with Stokes $I$, $Q$ and $U$. In addition to the brighter central disk, they saw ring-like pseudodisk surroundings around VLA 1623A. This surrounding extended envelope shows a spiraling structure in the pseudodisk, reminiscent of the spiraling inflows discussed our study, making VLA 1623A a potential candidate. \citet{Harris2018} repeated a similar observation but with polarized 872 $\mathrm{\mu}$m dust emission. They discovered that VLA 1623A contained two inner components, and that their polarization matched well the 1.3 mm observations. 
The polarization directions along the ring of \citet{Sadavoy2018} follow different pattern from ours, although \citet{Harris2018} deduce that observed polarization is likely a result of self-scattering and not of magnetic alignment. Therefore, polarization estimates are inconclusive in terms of our model. Observations focused on the  VLA 1623A ring/pseudodisk could illuminate the matter further.

For VLA 1623A, \citet{Murillo2013SMA} show that there is an outflow, with potentially two components, emerging from the object. While our \textit{LS} case produces an outflow, it is not continuous, which is in conflict with the observation. However, the LKS13 did not focus on modeling the outflow development. The Model H shows tentative hints that a more continuous, narrow outflow could be happening, but the dataset is too short to be conclusive. Therefore, extending the simulation beyond the current end point might address if a more stable outflow would emerge at a later stage. 
In both Model G and Model H, the velocity of the polar inflow is $\sim 0.4\kms$ in magnitude, which is supersonic, but slower than e.g. \citet{Bjerkeli2016} ($2.5$ -- $5.5\kms$) TMC1A or \citet{Murillo2013SMA} ($2$ -- $15\kms$) VLA 1623 outflows. This is approximately half of the free fall velocity with respect to the central point mass. 

Looking at VLA 1623A, \citet{Murillo2013} measured radial velocities from C$^{18}$O line emission. Their radial velocity measurements did not show direct spiral signatures, they found that their infall + Keplerian rotation model fit best to their observed PV-distribution, which is at least coherent with the picture presented in our study. Because of the observational challenges regarding polarization, more light could be shed on the surrounding ring by carefully examining velocities of the VLA 1623A pseudodisk, as our model would suggest spiraling variations in the velocity field. In such a way observing the velocities of VLA 1623A could also tell us something about its magnetic nature.

In the light of our results, the observations of IRAS 16293 by \citet{SadavoyIRAS2018} are also intriguing \citep[See also][for other polarization measurements]{Rao2009, Liu2018}. They performed their observations in ALMA Band 6 (1.3 mm) continuum with Stokes $I$, $Q$ and $U$, essentially following a comparable approach to \citet{Sadavoy2018}. However, using two separate pointings they got a measurement of a part of the extended emission between the IRAS 16293 A and B components. There they find polarization structures, which they call ``Streamers'' and a ``Bridge''. 

IRAS 16293 system is clearly more complicated than our spherically symmetric model. However, if we can believe that the polarization vectors around the edges of the IRAS 16293B disk trace magnetic fields, we have a case of magnetic alignment similar to what we see as spiralling inflow --- where the magnetic field is bending in rotational direction around towards the central disk. Therefore, there would be magnetic fields lines which outside of the disk align with rotational inflow analogous to our pseudodisk spirals. In such a case, the Bridge and B-Streamer of \citet{SadavoyIRAS2018} are good candidates for pseudospiral-type inflow.

\section{Conclusions}\label{sec:conclusions}

We have examined the time development of the MHD models of LKS13 using our radiative transfer code \pers. The MHD models were created to test the effects of misaligned (especially orthogonal) rotation axis and magnetic field direction during prestellar core collapse to solve the magnetic braking catastrophe. LKS13 found that misalignment could make disk formation possible if the mass to flux ratio was $\lambda \gtrsim 4$.
We have used \pers\ to continue the earlier analysis by showing how the examined systems would be visible in terms of column densities and Stokes $I$, $Q$ and $U$. We also analyzed how the velocity profile behaves in relation to what was seen and how the disk differs from simple circular Keplerian orbit. 
The 3D view of the simulation results and their time evolution has offered a number of insights not well visible from more traditional cut-slice approach. In a collapse model not all features are aligned with well determined planes or reach a genuine steady state. Therefore, those features will be missed by a slice cut or by a single snapshot of the intrinsically three-dimensional and time-evolving system.

We prepared a face-on view movie of CS Model G, showing column density together with streamlines.
From that movie we learn that the gas makes strongly eccentric and approximately elliptical orbits during its infall.
These infalling elliptical orbits constitute an inspiraling motion. The inner and outer turning points of that motion (roughly the periastron and apastron of the orbits) correspond to regions of increased column density; these dense regions largely match the ridges of (respectively) the inner and outer parts of the disk spirals as seen in the column density view.
These eccentric orbits are not exactly Keplerian; we know that kinematically because of their visible infall, and we know that dynamically due to the presence of a substantial self-gravity. Self-gravity is presumably the cause of the observed precession of the periastron (by making the orbits not closed), and visualizations show that it can exert torques. Magnetic forces are also known to be present in this system. Further study of this matter may require new simulations including (1) tracer fields helping to follow streaklines in addition to streamlines and (2) alleviating the disruptive effects of the numerical inner boundary conditions to the inner spiral when the periastron of the eccentric orbits becomes comparable with 6.7 AU.
For now we are ready to state that in addition to the disk spirals visualized and possibly detectable as a column density, the spirals visualized as streamlines allow us to say that the channel of motion is not necessarily exactly equal to the location of the spiral ridge path of highest densities, due to the presence of these eccentric orbital motions.

With these methods we improve our understanding of the MHD model and of its observable properties. In terms of the physical model we see:
\begin{enumerate}
    \item Collapse of a prestellar core with magnetic field perpendicular to the rotation axis leads to a spiral-like system. Formation of the disk is affected by snail shell like a pseudodisk, having projected appearance of a spiral. 
    
    \item The system is in the state of constant change and movement. Many of its essential features do not become visible without looking at their time evolution.
    
    \item Magnetic forces which can lead into visible precession of the disk during formation.

    \item Funnel-shaped axial infall flow (\textit{infunnel}) along polar directions is present during misaligned collapse.
    
    \item There is no visible continuous outflow at this early stage of misaligned collapse. If it happens it is momentary, bullet-like. 
    
    \item Spirals appear as velocity variations from circular Keplerian orbital velocity. Generally velocities are sub-Keplerian, especially in \textit{LS}-case, but in \textit{CS}-case super-Keplerian motions are possible in the inner disk along the spiral ridges. Some of the super-Keplerian effects are present due to eccentricity of disk rotation.  
\end{enumerate}

We found a number of potentially observable features connected to misaligned disk formation. The following features could provide a reasonable indicator of misaligned collapse:

\begin{enumerate}
    \item The general shapes of the rotationally supported disks are distinctively spiral-like, particularly with spiral-like pseudodisk in surroundings.
    
    \item Magnetically aligned dust polarization is dominated by the azimuthal direction of the magnetic field. Spiral characteristics can be found at the boundary of disk and pseudodisk. However, the actual spiral characteristics can be lost, with noise affecting polarization. Therefore, high sensitivity is required. 

    \item Observing spiral-like perturbations from velocity profiles with a Keplerian power law index could be the best way of recognizing the magnetic spirals. These perturbations should extend to the pseudo-disk. 

    \item The infunnels which appear during the collapse process could easily be mistaken for an outflow cavity if viewed sideways. However, as low density areas they might be too faint to be visible.
    
\end{enumerate}

With improved modeling efforts and as new results emerge from ALMA of early prestellar objects, misaligned disk model could be further tested. Misalignment of rotation axis and magnetic field present one aspect of disk formation, separate from from pure turbulent reconnection diffusion or non-ideal MHD effects. While the reality will likely be some combination of all of them, recognizing the type of influence of all such processes will help us to understand what is most important in which stage and context.

\acknowledgments

The authors thank prof.{} Ronald Taam for useful conversations during the research process. ZYL is supported in part by NSF grant AST-1815784 and AST-1716259 and NASA grant 80NSSC18K1095 and NNX14AB38G. KHL acknowledges support from an NRAO ALMA SOS award. This work utilized tools developed by the CHARMS group and computing resources and cluster in TIARA at Academia Sinica. This research has made use of SAO/NASA Astrophysics Data System.

\bibliographystyle{aasjournal}
\bibliography{references}

\appendix

\section{Velocity averages and gaussian fitting}

\label{sec:velgauss}

\subsection{Velocity averages}\label{sec:meanvel}

To see how the observed maps relate to apparent velocities in the system, we calculate the line of sight velocity averages for all Cartesian components $v_\mathrm{LOS}$, $v_\mathrm{W}$ and $v_\mathrm{N}$ --- or line of sight, west and north directional velocities respectively --- as a density weighted average
\begin{equation}
    \langle v_k \rangle = \frac{1}{\Sigma} \int \rho v_k \, ds. 
\end{equation}
Here $k$ denotes any velocity component. While only $v_\mathrm{LOS}$ would be observable in principle, all components can help with understanding the results in physical terms.

In this study, when streamlines are plotted or radial, poloidal and toroidal velocity component are discussed, they are calculated from the Cartesian velocity averages, assuming counterclockwise rotation of the system. Therefore, they are averages by projection, but not picked up directly from a cut from the three-dimensional model. This approach serves well when relating velocities to the also otherwise projected structures of column densities and Stokes parameters.

\subsection{Gaussian fitting}\label{sec:gauss}

We approached quantifying the time development with Gaussian fitting. For this, we focused on the edge-on observations (with the disk placed largely along the horizontal direction), as other viewing angles are not well suited for the method, which worked as follows.

First, we fitted a Gaussian profile to the Stokes I maps along the vertical direction for all horizontal points. This allowed us to get the local maximum intensity $I_{\mathrm{max,}i}$, distribution center $Z_i$ and the width of the disk $H_{z,i}$, as in 
\begin{equation}
    I_i(z) = I_{\mathrm{max},i}\exp\left(-\frac{(z-Z_i)^2}{H_{z,i}^2}\right),
\end{equation}
where $i$ is a pixel column along the horizontal axis. 

Based on the fit, we tried to estimate a number of disk properties and how they change by taking further averages. There are essentially the scale height $H_{z}$, disk tilt angle $\xi$ and peak locations $\zeta_{E}$ and $\zeta_{W}$. 

The height $H_{z}$ is calculated simply as a $I_{\mathrm{max},i}$ weighted average of $H_{z,i}$. The angle $\xi$ is derived be making a linear fit to $Z_i$ and calculating the relative angle of the fit, so that $\xi = 0\degree$ denotes the situation where the disk is perfectly horizontal. The average peak location $\zeta = (\zeta_{E} + \zeta_{W})/2$ is used to trace oscillations of the disk plane itself. The visible density of the disk is not horizontally uniform. There are local maxima in the intensities on the both sides of the sink particle/inner boundary (see e.g. Figure \ref{fig:GmaxI}). Therefore we trace the locations of such peaks, $\zeta_E$ on the east side $\zeta_W$ on the west side and trace their position over time, after taking an average, resulting in Figure \ref{fig:lombscargle} (top). Results of the Gaussian fitting analysis are presented in Section \ref{sec:ts}.

Apparent momentum $p_\mathrm{mom}$ was calculated by taking the average of $\Sigma | \langle v_\mathrm{LOS} \rangle |$ around the fit axis, or
\begin{equation}
    p_{\mathrm{mom},i} = \bigg\langle \Sigma | \langle v_\mathrm{LOS} \rangle | \bigg\rangle_{Z_i - H_{z,i}/2}^{Z_i+H_{z,i}/2}.
\end{equation}
While this is not an accurate measurement of the true momentum, it can function as a comparative measure between model results, functioning as a diagnostic variable when tracing the time evolution of motion.

\end{document}